\documentclass[%
 reprint,
 amsmath,amssymb,
 aps,
 prx,
]{revtex4-2}

\usepackage{graphicx}
\usepackage{dcolumn}
\usepackage{bm}
\usepackage{color}
\usepackage[colorlinks]{hyperref}
\usepackage{dsfont}
\usepackage{pst-node}%
\usepackage{array,mathtools,amssymb,booktabs}
\usepackage{color,comment}

\begin{document}

\title[]{Diverse communities behave like typical random ecosystems }

\author{Wenping Cui}
\email{cuiw@bc.edu}
\affiliation{Department of Physics, Boston University, 590 Commonwealth Avenue, Boston, MA 02139}
\affiliation{ Department of Physics, Boston College, 140 Commonwealth Ave, Chestnut Hill, MA 02467}
\author{Robert Marsland III}
\email{marsland@bu.edu}
\affiliation{Department of Physics, Boston University, 590 Commonwealth Avenue, Boston, MA 02139}

\author{Pankaj Mehta}%
\email{pankajm@bu.edu}
\affiliation{Department of Physics, Boston University, 590 Commonwealth Avenue, Boston, MA 02139}
\date{\today}
       
\begin{abstract}
In 1972, Robert May triggered a worldwide research program studying ecological communities using random matrix theory. Yet, it remains unclear if and when we can treat real communities as random ecosystems. Here, we draw on recent  progress in random matrix theory and statistical physics to extend May's approach to generalized consumer-resource models. We show that in diverse ecosystems adding even modest amounts of noise to  consumer preferences  results in a transition to ``typicality'' where macroscopic ecological properties of communities are indistinguishable from those of random ecosystems, even when resource preferences have prominent designed structures.  We test these ideas using numerical simulations on a wide variety of ecological models. Our work offers an explanation for the success of random consumer resource models in reproducing experimentally observed ecological patterns in microbial communities and highlights the difficulty of scaling up bottom-up approaches in synthetic ecology to diverse communities.
\end{abstract}

\maketitle


\section{Introduction}
One of the most stunning aspects of the natural world is the immense diversity of ecological communities ranging from rainforests to human microbiomes. Ecological communities are critical for numerous processes ranging from global water cycling processes\cite{spracklen2012observations} to animal development and host health\cite{belkaid2014role}. For this reason, understanding the principles governing community assembly and function in diverse communities has wide ranging applications from  conservation efforts to pharmaceutical engineering and bioremediation\cite{prosser2007role}. 

Many traditional ecological models focus on ecosystems consisting of a few species and resources. In such low dimensional models, it is often possible to characterize the ecological traits of all the species and resources and then use this information to make predictions about community-level properties \cite{friedman2017community, friedman2017ecological, ratzke2018ecological}. However, many natural communities are extremely diverse and the models and parameters are naturally high dimensional. This problem is especially pronounced in in the context of microbial ecology where hundreds of species can coexist in a single location. In this case, a comprehensive  parametrization of species and resource traits is no longer feasible, suggesting that new ideas and concepts are required to understand diverse communities.

A similar problem is encountered in statistical physics. For example, an ideal gas is characterized by the unit mole, which has the order of $10^{23}$ particles, making it impossible to simultaneously specify the microscopic state of the system (e.g. the positions and velocities of all particles). Despite this uncertainty, it is still possible to make predictions about macroscopic properties like pressure and the average energy by treating the positions and velocities of particles as independent random variables\cite{ma2018modern}. The fact that such universal statistical behaviors emerge naturally in large disorder systems composed on many particles suggests that a similar approach maybe possible in ecological systems.

\begin{figure*}
	\centering
	\includegraphics[width=0.8\textwidth]{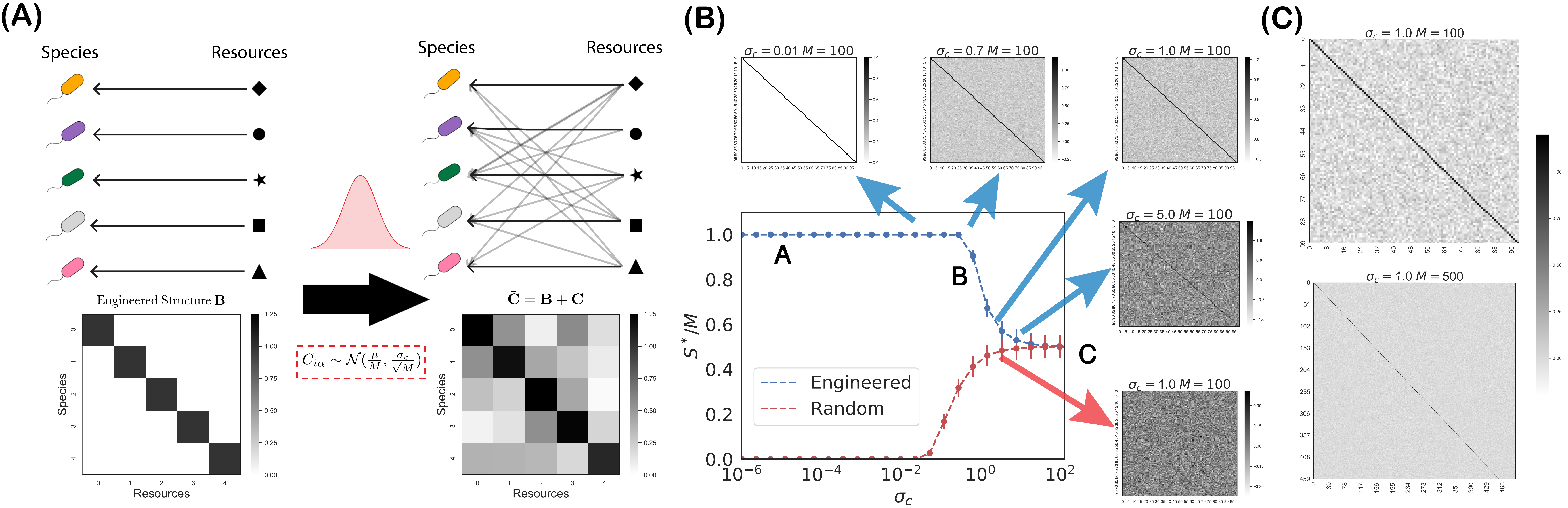}
	\caption{{\bf Random interactions destabilize an ecosystem of specialist consumers.} \textbf{(A)} Left: an ecosystem with system size $M=5$ starts with specialists consuming only one type of resource, resulting in a consumer preference matrix $\mathbf{B}=\mathds{1}$. Right: off-target consumption coefficients $\mathbf{C} \sim \mathcal{N}(\frac{\mu}{M}, \frac{\sigma_c}{\sqrt{M}})$ are sampled from a Gaussian distribution, resulting in an overall consumer preference matrix $\bar{\mathbf{C}}=\mathbf{B} +\mathbf{C}$. \textbf{(B)}  Fraction of surviving species $S^*/M$ vs. $\sigma_c$, numerically computed using $M = 100$ for an ecosystem described by Eq. \ref{Ma1}, along with the corresponding results for a completely random ecosystem with $\mathbf{B}=0$. The error bar shows $\pm 1$ standard deviation from 10000 independent realizations. Also shown are examples of the matrices $\bar{\mathbf{C}}$ employed in the simulations. \textbf{(C)} Heatmap for the identity matrix plus a gaussian random matrix with $\sigma_c=1$ for two system sizes: $M=100$ and $M=500$. }
	\label{heatmap}
\end{figure*}

In 1972, Robert May suggested that large complex ecosystems can also be modeled as random systems \cite{may1972will}. May considered a diverse ecosystem composed of $S$ species whose interspecific interactions were sampled randomly and independently from a normal distribution with zero mean and variance $\sigma^2$. In particular,  May asked when such a diverse random ecosystem would be stable to small perturbations. To answer this question, he examined the largest, i.e., the rightmost eigenvalue $\lambda_{\rm max}$ of the $S\times S$ community interaction matrix $\mathbf{J}$, whose diagonal entries -- chosen to be $J_{ii}=-1$ by May-- describe intraspecific competition and off-diagonal entries $J_{ij}$ describe how much the growth rate of species $i$ is affected by a small change in the population $N_j$ of other species $j$ from its equilibrium value. Using a mathematical formula for the distribution of eigenvalues of large random matrices derived by Ginibre \cite{ginibre1965statistical}, May showed that  $\lambda_{\rm max}$ increases with $S$, and derived a stability criterion governing the maximum diversity of an ecosystem: a diverse ecosystem becomes unstable to small perturbations when $\sqrt{S}\sigma>1$\cite{may1972will}. May's stability criteria has proven to be robust against a wide array of changes in the assumptions, including adding biologically realistic correlation structures to the matrix, or incorporating the dependence of the community matrix on population sizes in the Lotka-Volterra model \cite{gross2009generalized,allesina2012stability}.

In May's model, all ecosystem properties are encoded in the species-species interaction matrix. A major limitation of these models is that they neglect resource dynamics, making it difficult to understand how ecosystem properties depend on both the external environment and species consumer preferences. For this reason, community assembly is often analyzed using generalized Consumer Resource Models (CRMs)\cite{macarthur1967limiting,tilman1982resource}. In these models, species are modeled as consumer that can consume resources, and sometimes also produce resources  \cite{goldford2018emergent, marsland2020community, marsland2020minimal, tikhonov2017collective, posfai2017metabolic}. Recently, we have shown that  such models, initialized with random parameters, can predict lab experiments on complex microbial communities \cite{goldford2018emergent, marsland2020minimal} and reproduce large-scale ecological patterns observed in field surveys, including the Earth and Human Microbiome Projects \cite{marsland2020minimal}. This suggests that the large-scale, reproducible patterns we see across Microbiomes are emergent features of random ecosystems.

Yet, it remains unclear why random ecosystems can accurately describe real ecological communities. To answer these question, in this paper we exploit ideas from random matrix theory and statistical physics to analyze generalized consumer-resource models in spirit of May's original analysis. We show that the macroscopic ecological properties of diverse ecosystems can be described using random ecosystems, much like thermodynamic quantities like pressure and average energy of the ideal gas can be described by considering particles to be random and independent.

\subsection{Models}
To explore these ideas, we devised a more concrete version of May's original thought experiment describing an ecosystem consisting of $S$ non-interacting species where interactions are gradually turned on. May's original argument only considered the local dynamics near a pre-specified equilibrium point that eventually becomes unstable. Since we are interested in exploring what happens in consumer resource models, we must make additional modeling assumptions to arrive at a complete set of nonlinear dynamics. We focus on numerous variants of the Consumer Resource Model (CRM)\cite{macarthur1967limiting}, including different choices of resource dynamics, consumer preferences, as well as more dramatic variants such as the Microbial Consumer Resource Model introduced in \cite{goldford2018emergent, marsland2020community, marsland2020minimal}.

The original MacArthur Consumer Resource Model \cite{macarthur1967limiting} consists of $S$ species or consumers with abundances $N_i$ ($i = 1 ...S$) that can consume one of $M$ substitutable resources with abundances $R_\alpha$ ($\alpha = 1 ...M$), whose dynamics are described by the equations
\begin{eqnarray}\label{Ma}  
\begin{cases}
\frac{d N_{i} }{dt}  = N_{i}(\sum_{\beta} \bar{C}_{i \beta} R_{\beta} -  m_i)\\
\frac{d R_\alpha }{dt}    = R_\alpha(K_{\alpha}  - R_\alpha -  \sum_{j}N_{j} \bar{C}_{j\alpha}).
\end{cases}
\end{eqnarray}
The consumption rate of species $i$ for resource $\alpha$ is encoded by the entry $\bar{C}_{i\alpha}$ in the $S \times M$  consumer preference matrix $\mathbf{\bar{C}}$, $K_\alpha$ is the carrying capacity of resource $\alpha$, and $m_i$ is a maintenance energy that encodes the minimum amount of energy that a species $i$ must harvest from the environment  to survive. When the system is in the steady state, some species and resources can vanish. We denote the numbers of surviving  species and resources by $S^*$ and $M^*$, respectively, and in general at steady state we will have $S^*\le S$ and $M^* \le M$.  For this reason, we refer to this model as the CRM \emph{with} resource extinction and consider its effects analytically and numerically in Section \ref{resource_extinction} and Appendix \ref{subsec:withrd}. 

In the beginning, we  focus primarily on a popular variant of  the original CRM introduced by Tilman with slightly different resource dynamics\cite{tilman1982resource}:
\begin{eqnarray}\label{Ma1}  
\begin{cases}
\frac{d N_{i} }{dt}  = N_{i}(\sum_{\beta} \bar{C}_{i \beta} R_{\beta} -  m_i)\\
\frac{d R_\alpha }{dt}    = K_{\alpha}  - R_\alpha -  \sum_{j}N_{j} \bar{C}_{j\alpha}.
\end{cases}
\end{eqnarray}
From an ecological perspective, there are significant differences between this model variant and the original CRM. First, the resource supply rate $K_\alpha$ is constant instead of following logistic growth rate. Second, the species consume resources at a rate that is independent of the  resource concentrations in the environment. This can lead to unphysical, negative resource concentrations. Despite these differences, mathematically the equilibrium solutions of the two models have similar forms. One major difference that does arise is that in the dynamics described by Eq.~\ref{Ma1} consumers can no longer cause a resource to go extinct (i.e., $\mathbf{M^*}= \mathbf{M}$). This makes this models significantly easier to analyze (especially within the context of Random Matrix Theory) and leads to much simpler analytic expressions. For this reason, we largely focus on  this latter model \emph{without} resource extinction (see  Fig. \ref{distributions}, Fig. \ref{structures}, Fig. \ref{chi_si} for dynamics described by Eq~\ref{Ma1} and Appendix \ref{sec:additionfigures}  for numerics and  Appendix \ref{Cavity} for analytics on original CRM described by Eq. \ref{Ma}). Despite the unphysical, negative resource concentrations, the CRM without resource extinction captures almost all the qualitative behaviors present in more complicated and physically realistic CRMs (though there are some subtle but important differences discussed below).  

Both the models in Eq. \ref{Ma} and Eq. \ref{Ma1} make very specific assumptions about resource dynamics. To check the generality of our results, we also numerically analyzed generalizations of the CRM including linear resource dynamics where resources are supplied externally, and a model of microbial ecology with trophic feedbacks where organisms can feed each other via metabolic byproducts\cite{goldford2018emergent,  marsland2018available,marsland2020minimal, marsland2020community}. This analysis can be found in  Appendix \ref{sec:model}. Furthermore, for simplicity, in most of this work we assume that $S=M$. However, we have numerically checked that our results are robust to breaking on this assumption (see Fig. \ref{size}).

\begin{figure}
	\centering
	\includegraphics[width=0.5\textwidth]{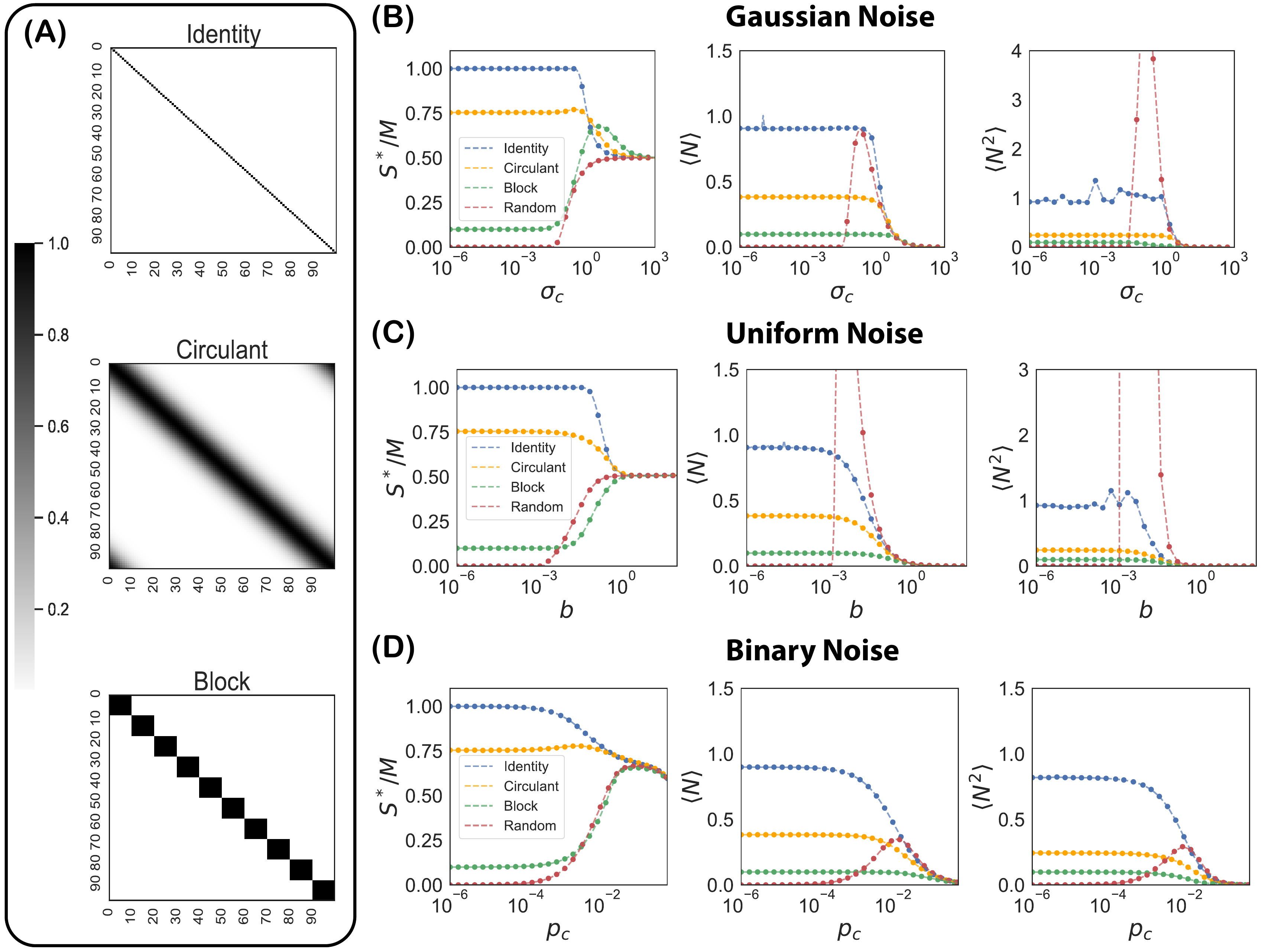}
	\caption{{\bf Community properties for structured and random ecosystems.} \textbf{(A)}: Examples of designed interactions   Top: the identity matrix; Middle: a Gaussian-type circulant matrix; Bottom: a block matrix (see Appendix \ref{sec:model} for details). Simulations of designed and random ecosystems where the random component of the the consumer preferences $\mathbf{C}$ are sampled from a \textbf{(B)}  Gaussian distribution $\mathcal{N}(0,\frac{\sigma_c}{\sqrt{M}})$,  \textbf{(C)} Uniform Distribution: $\mathcal{U}(0,b)$ or a \textbf{(D)}: Binomial distribution: ${Bernoulli}(p_c)$. The plots show the fraction of surviving species $S^*/M$, mean species abundance $\left< N \right>$, and second moment of the species abundances $\left< N^2 \right>$ for designed and purely random ecosystems ($\mathbf{B}=0$) the number of non-specific consumer preferences is increased.}
\label{phase}
\end{figure}

In CRMs, the identity of each species is specified by its consumption preferences. In real ecosystems, it is well established that organisms can exhibit strong consumer preferences for particular resources. However, recent work has shown that consumer resource models with random consumer preferences can reproduce experimental observations in field surveys and laboratory experiments \cite{goldford2018emergent, marsland2020minimal}. To understand this phenomena, we asked how  adding noise to consumer preferences changes macroscopic ecosystem level properties like diversity and average productivity. 
 To do so, we considered a thought experiment where we started with predesigned consumer resource preference, and then added ``noise'' to the consumer resource preferences. Mathematically, we can decompose the consumer matrix $\bar{\mathbf{C}}$ in Eqs \ref{Ma} and \ref{Ma1} into two parts: 
\begin{equation*}
\bar{\mathbf{C}}=\mathbf{B}+ \mathbf{C},
\end{equation*}
where  $\mathbf{B}$ encodes pre-designed structures, and  $\mathbf{C}$ is a random matrix representing "noise".

For simplicity, we started with non- interacting species where each species consumes its own resource.  A set of non-interacting species can be constructed by engineering each species to consume a different resource type, with no overlap between consumption preferences. For example,  one can imagine designing strains of \emph{E. coli} where each strain expresses transporters only for a single carbon source with all other transporters edited out of the genome: i.e a strain that can only transport lactose, another strain that can only transport sucrose, etc.  An ecosystem with such consumer preference structure is shown in Figure \ref{heatmap}(A). In such an experiment, horizontal gene transfer would eventually begin distributing transporter genes from one strain to another, so a realistic model would have to allow for some amount of unintended, ``off-target'' resource consumption. In line with May, we can model the consumer  preferences $\bar{C}_{i\alpha}$ of species $i$ for resource $\alpha$  in such an ecosystem as the sum of the identity matrix $\mathbf{B}=\mathds{1}$ and a random component $C_{i\alpha}$ with variance $\sigma^2$ that encodes non-specific preferences (see Figure \ref{heatmap}A right). In other words, the full consumer matrix can be written as $\bar{\mathbf{C}} =\mathbf{I} +\mathbf{C}$.

\section{Results}
\subsection{Phase transition to random ecosystems}
Figure \ref{heatmap}(B) shows how the number of surviving species at steady-state changes as one adds more and more non-specific resource preferences to an ecosystem initially composed of non-interacting species. Just as in May's analysis, the appropriate measure of the importance of the random component is the root-mean-squared off-target consumption $\sigma_c = \sqrt{M\sigma^2}$ (recall $M=S$). This scaling reflects the fact that two consumer matrices $\bar{\mathbf{C}}$ with the same $\sigma_c$ but different system sizes $M$ can have very different amounts of absolute noise as shown Figure \ref{heatmap}(C), but exhibit almost identical community-level properties (with all differences coming from finite size effects, see Fig. \ref{Msize} for the universal behavior at different $M$).  Figure \ref{heatmap}(B)  shows the fraction of surviving species $S^*/M$ in the ecosystem as a function of $\sigma_c$. At small values of $\sigma_c$, all the species survive and $S^*=S$. As high as $\sigma_c = 0.7$, almost all of the original species are still present in the community. But between $\sigma_c = 0.7$ and $\sigma_c = 1$, there is a sharp transition in community structure, which results in about half of the original species becoming extinct. 

Remarkably, the fraction of surviving species converges to the same value as for a completely random consumer preference matrix and remains finite as $\sigma_c \to \infty$ \cite{servan2018coexistence}. This means that ecosystems with an arbitrarily large number of species can be stably formed by considering a sufficiently large initial species pool. We also examined two other community-level properties: the mean species abundance $\langle N \rangle$ (i.e., the average productivity), and the second moment of the population size $\langle N^2 \rangle$, which includes information about the distribution of population sizes of various species. Figure \ref{phase} shows that both of these quantities are also well-approximated by the random consumer preference matrix for $\sigma_c > 1$. These numerical predictions are in excellent agreement with analytic predictions derived in the $S\to\infty$ limit derived in  Appendix \ref{Cavity} using the cavity method \cite{bunin2017ecological,advani2018statistical}.

This convergence to random ecosystem behavior is quite robust, and holds for other choices of designed consumer preferences beyond the identity matrix considered above. Figure \ref{phase} shows numerical simulations of the diversity $S^*/M$,  average productivity $\langle N \rangle$, and second moment of the species abundances $\langle N^2 \rangle$ as a function of the noise $\sigma_c$  for two other choices of designed consumer preference matrices: a block structure with pre-defined groups of species exhibiting strong intra-group competition and a unimodal structure where each species is more likely to consume resources similar to its preferred resource. Once again, we see that the ecosystem quickly transitions to a behavior where these macroscopic properties are indistinguishable from those of a random ecosystem. Borrowing terminology from physics, we call systems whose macroscopic properties are well described by random ecosystems as \emph{typical}. The primary effect of the choice of consumer preference matrix is to adjust the threshold value of $\sigma_c$ where the transition to typicality occurs. In all cases, we find that the random behavior takes over when the average total off-target consumption capacity over all $M$ resource types becomes greater than the consumption of the primary resource in the original designed ecosystem in the absence of noise.

The character of the self-organized state is also robust to changes in the sampling scheme for the random component of the consumer preferences. Gaussian noise in consumer preferences simplifies the analytic calculations but also sometimes results in non-physical negative values for consumer preferences. We therefore tested two sampling schemes that always produce positive values for consumer preferences: uniformly sampling the random component of preferences $C_{i \alpha}$ in an interval from 0 to $b$, and binary sampling where $C_{i \alpha}=1$ with probability $p_c$ and zero otherwise. Changing $b$ or $p_c$ affects both the mean and the variance of the random components of the consumer preferences simultaneously making it difficult to directly compare to the Gaussian case. Nonetheless, as can be seen in the Figure \ref{phase}, the qualitative behaviors is identical to the Gaussian case, with macroscopic ecological properties becoming indistinguishable from those of a fully random ecosystem when the average off-target resource consumption comparable to the the consumption of the designed resources.
\begin{figure}
	\centering
	\includegraphics[width=0.5\textwidth]{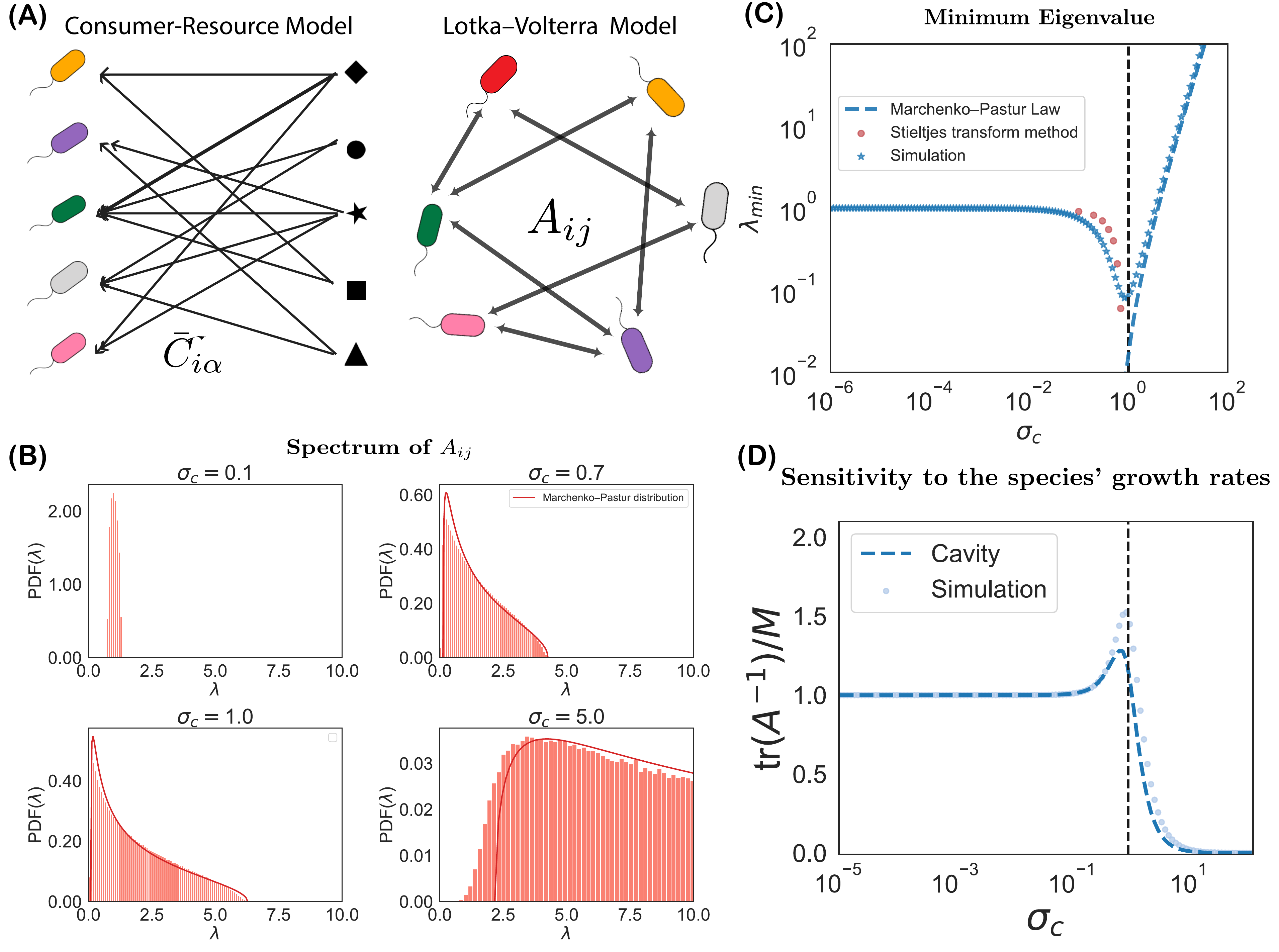}
	\caption{{\bf Effect of random interactions on ecosystem sensitivity}. \textbf{(A)}: The bipartite interactions $\bar{C}_{i\alpha}$ in MacArthur's consumer-resource model can be mapped to pairwise competition coefficients $A_{ij}$ in generalized Lotka-Volterra equations through $A_{ij}=\sum_{\alpha\in \mathbf{M}}\bar{C}_{i\alpha}\bar{C}^{T}_{\alpha j}$. \textbf{(B)} Spectra of $A_{ij}$ at different $\sigma_c$ for $\bar{\mathbf{C}}=\mathds{1}+ \mathbf{C}$, where $\mathbf{C}$ is a random matrix with i.i.d entries drawn from a normal distribution with mean zero and standard deviation $\sigma_c$. The red solid line is the Marchenko-Pastur distribution. \textbf{(C)}: Comparison between numerical simulations and analytic results for the minimum eigenvalue of $\mathbf{A}$ at different $\sigma_c$. \textbf{(D)}: Comparison between numerical simulations and analytic solutions for the mean sensitivity $\nu$ of steady-state population sizes to changes in species growth rates. See Appendix \ref{SI:simulations} for details.}
\label{pdf}
\end{figure}

\subsection{Sensitivity to perturbations and the transition to typicality}
To better understand why mass extinctions happen at $\sigma_c^*\sim1$ and allow for comparison with May's original analysis, we calculated an effective species-species competition matrix $A_{ij}$ between species for an ecosystem whose dynamics are governed by  Eq. \ref{Ma1}.  We exploited the observation by MacArthur and others that if resource abundances always remain close to their steady state values, the steady-states of the CRM coincide with those of an effective generalized Lotka-Volterra model of the form 
\begin{eqnarray}\label{LV-PM}  
\frac{dN_i}{dt}=N_i\left (\sum_{\alpha\in \mathbf{M}} C_{i \alpha}K_\alpha-m_i- \sum_j A_{ij}N_j \right),
\end{eqnarray}
with the species-species interaction matrix given by
\begin{eqnarray}\label{Aij-PM}  
A_{ij}=\sum_{\alpha\in \mathbf{M}}\bar{C}_{i\alpha}\bar{C}^{T}_{\alpha j}
\end{eqnarray}
(see  Figure \ref{pdf}(A) and Appendix \ref{sec:mapLV} for details). This matrix is related to May's community matrix governing stability $\mathbf{J}$ discussed in the introduction through the relation $J_{ij} = -\bar{N}_i A_{ij}$, where $\bar{N}_i$ is the steady-state abundance of species $i$. For symmetric interaction matrices of the form in Eq. \ref{Aij-PM}, it is possible to prove that the largest eigenvalue $\lambda_{\rm max}$ of $\mathbf{J}$ reaches zero from below only when the smallest eigenvalue $\lambda_{\rm min}$ of $\mathbf{A}$ reaches zero from above (see Appendix \ref{sec:Jacobimatrix}). 

As shown in Figure \ref{heatmap}(B), the behavior broadly falls into one of three different regimes depending on the 
amount of noise introduced in the consumer preferences: a low-noise regime when  $\sigma_c \ll 1$, a  cross-over regime when  $0\ll\sigma_c \leq 1$, and a high-noise regime when $\sigma_c > 1$. Figure \ref{pdf}(B) shows how the eigenvalue spectrum of the
corresponding Lotka-Volterra  interaction matrix $\mathbf{A}$ change as $\sigma_c$ increases.

 \textbf{Low-noise regime ($\sigma_c \ll 1$):} 
In the low-noise regime, the engineered structure in the consumer preference controls large scale ecological properties. Furthermore, the eigenvalue spectrum of the LV-interaction matrix $\mathbf{A}$ is centered around $1$ reflecting the fact there is very little competition between species (i.e., species still occupy largely independent niches). For this reason, in this regime all the initial species in the ecosystem survive to steady-state so that ${S^*}/{M}=1$.

\textbf{Crossover regime ($0\ll\sigma_c \leq 1$):}
With increasing $\sigma_c$, the eigenvalues due the noise component in $\mathbf{A}$ repel each other like in the Coulomb gas and the spectrum spreads out \cite{dyson1962statistical}. $\lambda_{\rm min}$ decreases until it reaches the threshold of stability $\lambda_{\rm min} \cong 0$ at $\sigma_c^* \approx 1$. Note that $\lambda_{\rm min}$ is close to 0 but not exactly at 0 because the steady-state of the CRM is always stable \cite{chesson1990macarthur}. In this regime even a small environmental perturbations or small amounts of demographic noise can result in species extinctions \cite{dalmedigos2020dynamical}. This is closely related to the divergence of structural stability when $\lambda_{\rm min}\sim 0 $\cite{rohr2014structural}.  In Appendix \ref{Cavity} we show analytically using the Cavity method \cite{bunin2017ecological, advani2018statistical} that in the limit $M\to\infty$, $\lambda_{\rm min}$ is approaches 0 from above when $\sigma^*_c = 1$. At $\sigma_c \sim1$ the engineered structure and noise have comparable amplitudes. For the case where the consumer preferences are chose to be binary noise, this threshold corresponds to a critical noise level $p_c\sim \frac{1}{M}$, meaning on average there is one random nonzero element in the row besides the diagonal one. More generally, our numerics suggest that the threshold to typicality occurs in a wide variety of models when the expected off-target resource consumption rates become comparable to the consumption rate for the designed resources.


\textbf{Noise-dominated regime ($\sigma_c > 1$)}
In this regime, we observe two new phenomena that were not accessible in May's original framework. First, the spectrum of the species-species interaction matrix $A_{ij}$ approaches the Marchenko-Pastur law \cite{marchenko1967distribution}, 
\begin{equation}\label{MPlaw}
\resizebox{.43\textwidth}{!}{$\rho(x)\!=\!\frac{1}{2\pi\sigma_c^2 c x}\sqrt{(b-x)(x-a)}\!+\!\Theta(c-1)(1-c^{-1})\delta(x)$}
\end{equation}
where $a=\sigma_c^2(1-\sqrt{c})^2$, $b=\sigma_c^2(1+\sqrt{c})^2$,  $c={S^*}/{M}$ and $\Theta(x)$ represents the Heaviside step function. This differs from May's analysis where the spectrum of the interaction network  follows Girko's Circular law \cite{rogers2008cavity, altieri2019constraint, agliari2019marchenko}.
The reason for this difference is that species-species interaction matrix obtained from the CRM  is the outer product of a random matrix $\mathbf{\bar{C}}$ with itself (i.e., a Wishart matrix, see Eq. \ref{Aij-PM}), reflecting the fact that the CRM has two different kinds of degrees of freedom: resources and species. The Marchenko-Pastur law is the distribution we would expect for an ecosystem with completely random consumer preferences \cite{marchenko1967distribution}. This helps explain our earlier observations that community-level observables of ecosystems are indistinguishable from the purely random ecosystems when $\sigma_c$ is sufficiently large (see Figure \ref{pdf}(B)).

Secondly, as $\sigma_c$ increases past 1 and ecosystem properties become typical, the resulting ecosystems once again become insensitive to external perturbation \cite{dalmedigos2020dynamical}.  To see this, we note that we can measure sensitivity to perturbations by examining the minimum eigenvalue of the interaction matrix $A_{ij}$, with larger $\lambda_{\rm min}$ meaning decreased sensitivity to perturabations (see Appendix \ref{sec:Jacobimatrix}). The minimum eigenvalue in the Marchenko-Pastur Distribution is located at
\begin{equation}\label{MPlawsi}
\lambda_{\rm min} = \sigma_c^2(1-\sqrt{S^*/M})^2.
\end{equation}
As one increases $\sigma_c$,  $S^*/M \rightarrow 1/2$ from above since there is increases competition between species for shared resources. Consequently, $\lambda_{\rm min}$ is always  much larger than zero once ecosystems crossover to their typical behavior.

The above analysis suggests that  $\lambda_{\rm min}$ is an important property that can be used to characterize the three regimes seen in Figure \ref{pdf}(C).   In the low-noise regime, species-species interactions are weak and $\lambda_{\rm min} \approx 1$, whereas in the high-noise regime $\lambda_{\rm min} =\sigma_c^2(1-\sqrt{S^*/M})^2$. The calculation of $\lambda_{\rm min}$ in Regime B is challenging because of the mixture between the engineered structure and noise. However, we can use techniques from RMT for wireless communication (i.e information-plus-noise models) to analytically estimate $\lambda_{\rm min}$ \cite{couillet2011random, loubaton2011almost}. The results are shown in the red scatter points in Figure \ref{pdf}(D) (see Appendix \ref{ipnm}). As discussed above, $\lambda_{\rm min}$ approaches zero as $\sigma_c$ approaches one. 

The spectrum of $\mathbf{A}$ also contains quantitative information about the sensitivity of the ecosystem in the Cavity method. Specifically, as shown in Appendix \ref{Cavity}, we can define a susceptibility $\nu$ that measures the average response of the steady-state population size $\bar{N}_i$ to perturbing of the species maintenance cost $m_i$ (see Eq. \ref{Ma1}). We further show that  $\nu$  is directly related to the the sum of the inverse eigenvalues of $A_{ij}$ through the expression
\begin{equation}
\nu=\frac{1}{M}\sum_i (1/\lambda_i) = \frac{1}{M}{\rm tr}(\mathbf{A}^{-1}).
\end{equation}
Figure \ref{pdf}(D) shows that this quantity is initially constant as $\sigma_c$ is increased from 0, then reaches the maximum value at $\sigma_c = 1$, and finally rapidly decreases to near zero. In Appendix \ref{Cavity} we provide analytical calculations based on the cavity method confirming these numerical results. 

Note that our results are not restricted to Gaussian noise  but also apply to the other cases where the noise in consumer preferences is  binary or uniform  (See Figure \ref{matrix} and \ref{distributions}). This is because the \textit{central limit theorem} guarantees that  the statistics of eigenvalues of large random matrices converge to the statistics in Gaussian random matrices for many biologically plausible choices of consumer preferences. 

\begin{figure*}
	\centering
	\includegraphics[width=0.8\textwidth]{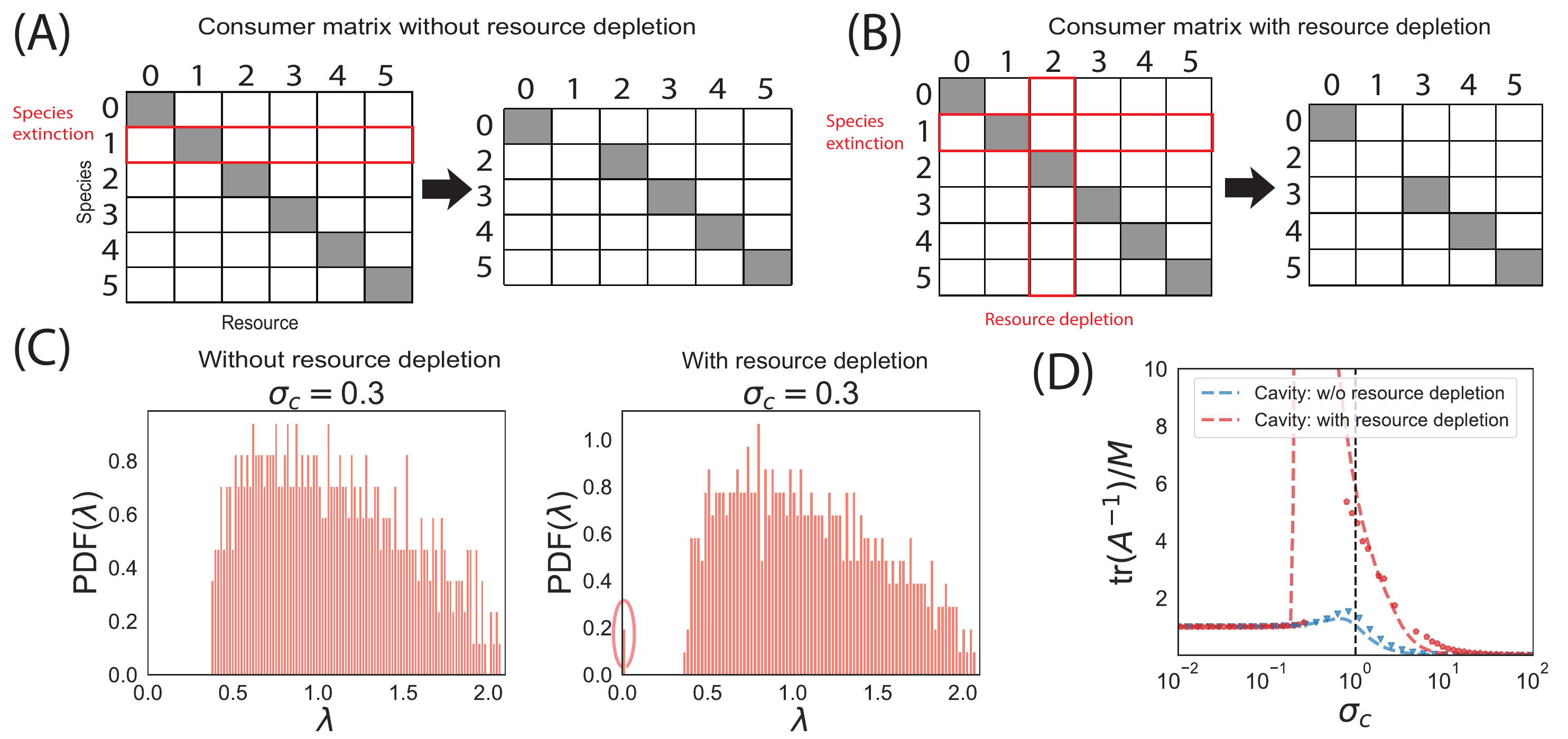}
	\caption{{\bf Effect of resource extinction on an ecosystem}. A schematic for the consumer preference matrix with (\textbf{(A)}) and (\textbf{(B)})without resource extinction for specialist consumers that each eat independent resources. The left schematic corresponds to the initial consumer matrix, and the right schematic to the consumer matrix after species and resource extinctions. Notice that resource extinctions can result in singular consumer matrices \textbf{(C)} Spectra of $A_{ij}$   at $\sigma_c=0.3$ with consumer matrices chosen as in Figure \ref{pdf} with (left) and without resource extinction (right). The zero modes are marked with a red ellipse. \textbf{(D)} the mean sensitivity $\nu$ of steady-state at different $\sigma_c$. The dashed lines in  \textbf{(D)} are cavity solutions. The scatter points are results from numerical simulations. See Appendix \ref{Cavity} for detailed calculations. }
	\label{rdelete}
\end{figure*}

\subsection{Effect of resource extinction}\label{resource_extinction}
Thus far we have focused on a CRM without resource extinctions specified by Eqs. \ref{Ma1}.  As discussed extensively in  Appendix \ref{Cavity}, if we  allow for resource extinction (Eqs. \ref{Ma}) and write  
\begin{equation}
A_{ij}=\sum_{\alpha\in \mathbf{M^*}}\bar{C}_{i\alpha}\bar{C}^{T}_{\alpha j}
\end{equation} 
instead of eq. \ref{Aij-PM}, somewhat surprisingly, our cavity method predicts a second-order phase transition to typicality rather than a cross-over as is the case without resource extinction. The signature of such a second order transition is the divergence of the susceptibility matrix $\nu$ discussed above. Figure \ref{rdelete} shows $\nu$ with and without resource extinction, numerically confirming the existence of this second order transition. This second order transition is also reflected in
the spectrum of the interaction matrix $\mathbf{A}$ through the the appearance of zero eigenvalue modes for CRMs when resources can go extinct.

The existence of zero modes can be understood by noting that resource extinction and species extinction correspond to the column and row deletion in the consumption matrix (shown in Figure \ref{rdelete}(A)). Such deletions can change the engineered component of the effective consumer preferences for surviving species and resources, resulting in large fluctuations
in the interaction matrix $\mathbf{A}$.  In the presence of these large fluctuations, the interaction matrix no longer self-averages, giving rise to the observed second-order phase transition. This same mechanism also leads to a second-order phase transition to typical behavior when the engineered portion of the consumer resources is block diagonal, even in the absence of resource extinctions (see Figure \ref{structures}).



\section{Discussion}

It is common practice in theoretical ecology to model ecosystems using random matrices. Yet it remains unclear if and when we can treat real communities as random ecosystems. Here, we investigated this question by generalizing May's analysis to consumer resource models and asking when the macroscopic, community level properties can be accurately predicted using random parameters. We found that introducing even modest amount of stochasticity into consumer preferences ensures that the macroscopic properties of diverse ecosystems will be indistinguishable from those of a completely random ecosystem.  Our calculations and numerics suggest that transition to typicality occurs when the total amount of off-target resource consumption becomes comparable to the consumption rate of targeted resources.

We confirmed our analytic calculations using numerical simulations on CRMs with different types of resource dynamics and different classes of non-specific interactions. We emphasize that despite the fact that random ecosystems can make accurate predictions about macroscopic properties like the average diversity or productivity, they will in general fail to capture species level details. This phenomena is well understood in the context of statistical physics where it is possible to predict thermodynamic quantities such as pressure and temperature even though one cannot accurately predict microstates. 

These observations may help explain the surprising success of consumer resource models with random parameters in predicting the behavior of microbial ecosystems in the lab and natural environments \cite{goldford2018emergent,marsland2020minimal}. They also suggest that maybe possible to predict macroscopic ecosystem level properties like diversity or total biomass even when ecosystems are poorly characterized or have lots of missing data.

The foregoing analysis has several other interesting implications. First, it suggests that bottom-up engineering of complex ecosystems may prove to be very difficult. As the number of components increases, small uncertainties in each of the interaction parameters may eventually overwhelm the designed interactions, and destabilize the intended steady state.  Instead, such  system are much more likely end up in a typical state which our theory suggests is much more stable than the intended designed state as ecosystems become more diverse.

Our work also suggests that in ecosystems well described by consumer resource models, crossing a May-like transition generically gives rise to typical random ecosystems rather than a marginal stable phase as was found in a recent analysis of the Generalized Lotka-Volterra model \cite{biroli2018marginally, roy2019numerical, altieri2020properties} (an important caveat to this statement is that adding non-resource based interactions to consumer resource models can restore complicated behavior reminiscent of the marginally stable phase \cite{dalmedigos2020dynamical}). This suggests that even when cumulative parameter uncertainties preclude a detailed characterization of an ecosystem, methods from statistical physics and Random Matrix Theory can be employed to predict system-level properties \cite{barbier2018generic, servan2018coexistence}.  It will be interesting to explore if and how these insights can be exploited to design  top-down control strategies for ecosystems and identify assembly rules for microbial communities with many species \cite{friedman2017community}.

In this paper, we only consider white noise, which is independently and identically added to all interaction components. In the future, it will be interesting to ask how
other specialized noises, resulting from demographic stochasticity,  phenotypic variation, can affect our results. Based on our experience, we expect that, even in these more complicated
ecosystems, our conclusion will hold quite generally in the thermodynamics limit. But much more work needs to be done to
confirm if this intuition is really correct.

\begin{acknowledgments}
	We thank Josh Goldford, Zhenyu Liao, Jason Rocks, Guangwei Si, Jean Vila and Yu Hu for helpful discussions. We also especially appreciate numerous valuable comments from Stefano Allesina. The work was supported by NIH NIGMS grant 1R35GM119461, Simons Investigator in the Mathematical
	Modeling of Living Systems (MMLS).
	The authors are pleased to acknowledge
	that the computational work reported on in
	this paper was performed on the Shared Computing Cluster
	which is administered by Boston University Research
	Computing Services.
\end{acknowledgments}
\bibliography{ref.bib}
\onecolumngrid

\appendix
\newpage
\pagebreak


\renewcommand{\thefigure}{A\arabic{figure}}
\setcounter{figure}{0}

\section{Model setup}\label{sec:model}
We primarily analyze CRMs of the form given by Eqs \ref{Ma} and \ref{Ma1}.  To do so, we decompose the consumer matrix $\bar{\mathbf{C}}$ into two parts: 
\begin{equation}
\bar{\mathbf{C}}=\mathbf{B}+ \mathbf{C},
\label{SI-CP}
\end{equation}
 with  $\mathbf{B}$ encoding a pre-designed set of resource-mediated interactions, and  $\mathbf{C}$ a random matrix encoding ``off-target'' consumption.  We  consider three types of  $\mathbf{B}$ (see Figure \ref{phase}):  the identity matrix, a square Gaussian-type circulant matrix  $B_{i\alpha}=e^{-\text{min}(i, |M-i|)^2/r^2}$ with  $r=7$\cite{butler2018stability} and a block matrix with identical  $10\times 10$ blocks(all elements are 1 inside the $10\times 10$ block). 
We also consider three types of random matrices $\mathbf{C}$. In all cases, each element in the matrix is sampled independently from an underlying probability distribution. The three distributions we consider are a normal distribution with mean zero and standard deviation $\sigma_c/\sqrt{M}$, a uniform distribution where each element is sampled uniformly from $[0, b]$, and  a Bernoulli distribution where each element can be $+1$ with probability $p_c$  and $0$ with probability $1-p_c$ (i.e Binary Noise).

For all simulations, unless otherwise specified the default choices for parameters are: $M=100$, $\mu=0$, $K=1$, $\sigma_K=0.1$ ,  $m=0.1$ and $\sigma_m=0.01$ and each data point is averaged from 5000 independent realizations. The simulation detail for each figure can be found at Appendix \ref{SI:simulations}.  All simulations are available on GitHub at \url{https://github.com/Emergent-Behaviors-in-Biology/typical-random-ecosystems}.

\subsection{Alternative Models used in Appendix}
To test the generality of our results, we also simulated more complicated variants of the consumer resource model (see  Figure \ref{dynamics} and Appendix \ref{linearSusceptibilities}). First, we simulated a consumer resource 
model with linear resource dynamics  \cite{cui2019effect}:
\begin{eqnarray}\label{Ma2}  
\begin{cases}
\frac{d N_{i} }{dt}  = N_{i}(\sum_{\beta} \bar{C}_{i \beta} R_{\beta} -  m_i)\\
\frac{d R_\alpha }{dt}    = \kappa_{\alpha}  - R_\alpha -  \sum_{j}N_{j} \bar{C}_{j\alpha}R_\alpha.
\end{cases}
\end{eqnarray}
In this model resources are supplied externally at a rate rather than described by a logistic growth.
This small change  in resource dynamics can significantly change the ecosystem properties because it prevents resources from going extinct in the steady state. In the simulations, we set $M=100$, $\mu=1$, $\kappa=1$, $\sigma_\kappa=0.1$ ,  $m=0.1$ and $\sigma_m=0.01$ and each data point is averaged from 1000 independent realizations. 

Second, we simulated a generalization of the MacArthur's Consumer Resource model to a model we call the Microbial Consumer Resource Model (MicroCRM). The MicroCRM
was introduced in \cite{goldford2018emergent} and refined in \cite{marsland2018available} to simulate microbial communities. In this model, in addition to consuming resources species can produce new resources
through cross-feeding. This dramatically changes the resource dynamics through the introduction of trophic feedbacks. Unlike the original CRM and the CRM with linear resource dynamics, the MicroCRM possesses no Lyapunov function. Full details of the model are available in the appendix of  \cite{marsland2018available, marsland2020community}.  In particular, the dynamics we use are described in equation (17) of \cite{marsland2018available} with the leakage rate $l=0.4$.  The fraction of secretion flux secreted to the same resource type is $f_s = 0.45$, the fraction of secretion flux to 'waste' resource is $f_w = 0.45$ and variability in secretion fluxes among resources is $d_0=0.2$. We set $M=100$, $\mu=1$, $K=1$, $\sigma_K=0.1$ ,  $m=0.1$ and $\sigma_m=0.01$ and each data point is averaged from 1000 independent realizations.

\section{Sensitivity to Parameter Perturbations}\label{Susceptibilities}
 We begin by defining four susceptibility matrices that measure how the steady-state resource and species abundances
 respond to changes in the resource supply and species death(growth) rates:
\begin{eqnarray}
\chi^R_{\alpha \beta}=\frac{\partial \bar{R}_\alpha}{\partial K_\beta},
\chi^N_{i \alpha}=\frac{\partial \bar{N}_i}{\partial K_\alpha}, 
\nu^R_{\alpha i}=\frac{\partial \bar{R}_\alpha }{\partial m_i}, 
\nu^N_{i j}=\frac{\partial \bar{N}_i }{\partial m_j}
\end{eqnarray}
where the bar $\bar{X}$ over the variable $X$  denotes the steady-state (equilibrium) solution.

For the extinct species and resources, by definition the susceptibilities are zero. For this reason, we focus only on the surviving resources and species.
At steady-state, equation (\ref{Ma}) gives:
\begin{eqnarray}
0&=&\sum_{\alpha\in \mathbf{M^*}} \bar{C}_{i \alpha} \bar{R}_{ \alpha} - m_i\\
0&=&K_{\alpha} - \bar{R}_\alpha- \sum_{j\in \mathbf{S^*}}\bar{N}_{j} \bar{C}_{j\alpha}
\end{eqnarray}
where $\mathbf{M^*}$  and $\mathbf{S^*}$ denote the sets of resources and species, respectively, that survive in the ecosystem at steady-state. Differentiating
these equations yields the relations
\begin{eqnarray} 
0\!=\! \sum_{\alpha\in \mathbf{M^*}} \!\bar{C}_{i \alpha} \frac{\partial \bar{R}_\alpha}{\partial K_\beta},  &\quad& \delta_{\alpha \beta}\!=\!\frac{\partial  \bar{R}_\alpha}{\partial K_\beta}\!+\! \sum_{j\in \mathbf{S^*}}\frac{\partial  \bar{N}_j}{\partial K_\beta} \bar{C}_{j\alpha} \nonumber \\
\delta_{i j}\!=\! \sum_{\alpha\in \mathbf{M^*}}\! \bar{C}_{i \alpha} \frac{\partial  \bar{R}_\alpha}{\partial m_j},  &\quad&
0\!=\!\frac{\partial  \bar{R}_\alpha}{\partial m_i}\!+ \!\sum_{j\in \mathbf{S^*}}\frac{\partial  \bar{N}_j}{\partial m_i} \bar{C}_{j\alpha}.
\end{eqnarray} 
Substituting in for the partial derivatives using the susceptibility matrices defined above, we have:
\begin{eqnarray} 
0=\sum_{\alpha\in \mathbf{M^*}} \bar{C}_{i \alpha} \chi^R_{\alpha \beta},&\quad&
\delta_{\alpha \beta}=\chi^R_{\alpha \beta}+ \sum_{j\in \mathbf{S^*}}\chi^N_{j \beta} \bar{C}_{j\alpha} \nonumber \\
\delta_{i j}=\sum_{\alpha\in \mathbf{M^*}} \bar{C}_{i \alpha} \nu^R_{\alpha j}, &\quad&
0= \nu^R_{\alpha i}+ \sum_{j\in \mathbf{S^*}} \nu^N_{j i} \bar{C}_{j\alpha}.
\end{eqnarray} 
These two equations can be written as single matrix equation for block matrices:
\begin{eqnarray}
\begin{pmatrix}
\bar{\mathbf{C}}& 0\\\
\mathds{1}& \bar{\mathbf{C}}^T
\end{pmatrix}  
\begin{pmatrix}
\mathbf{\nu}^R & \mathbf{\chi}^R\\\
\mathbf{\nu}^N& \mathbf{\chi}^N
\end{pmatrix} =\mathds{1}
\end{eqnarray}
To solve this equation, we define a $S^* \times S^* $ matrix: $A_{ij}=\sum_{\alpha\in M^*}\bar{C}_{i\alpha}\bar{C}^{T}_{\alpha j}$. A straightforward calculation yields
\begin{eqnarray}
\chi^R_{\alpha \beta}&=&\delta_{\alpha\beta}-\sum_{i\in \mathbf{S^*}}\sum_{j\in \mathbf{S^*}} \bar{C}^T_{\alpha i}A^{-1}_{ij}\bar{C}_{j\beta} \\
\chi^N_{i \alpha}&=&\sum_{j\in \mathbf{S^*}}A^{-1}_{ij}\bar{C}_{j\beta}\label{chi1}, \quad \nu^R_{\alpha i}=\sum_{j\in \mathbf{S^*}} \bar{C}^T_{\alpha j}A^{-1}_{ji } \\
\nu^N_{i j}&=&-A^{-1}_{ij},\quad  i,j\in \mathbf{S}^* \text{ and } \alpha,\beta\in \mathbf{M}^* \label{nu1}
\end{eqnarray}
\section{Cavity Solution}\label{Cavity}
When the designed component of the consumer preferences is the identity (i.e $\mathbf{B}=\mathds{1}$ in Eq.  \ref{SI-CP}), the effect of random off-target consumption on system-scale properties can be computed analytically in the $M,S \to \infty$ limit using the cavity method \cite{bunin2017ecological,advani2018statistical}. The cavity calculation is straightforward but tedious. For this reason, it is helpful to introduce the notation:
\begin{itemize}
	\item $\frac{M^*}{M}=\phi_R$, $\left< R\right>=\frac{1}{M}\sum_\beta R_\beta$ and $q_R=\frac{1}{M}\sum_\beta R^2_\beta=\left< R^2\right> $ , where $M^*$ is the number of surviving resources. 
	\item $\frac{S^*}{S}=\phi_N$, $\left< N\right>=\frac{1}{S}\sum_j N_j$ and $q_N=\frac{1}{S}\sum_j N^2_j=\left< N^2\right>$, where $S^*$ is the number of surviving species. 
	\item $C_{i \alpha} \equiv {\mu \over M} + \sigma_c d_{i \alpha}$ assuming $\left< d_{i \alpha}\right>=0$, $\left<d_{i \alpha} d_{j \beta} \right>={\delta_{ij} \delta_{\alpha \beta} \over M}.$  with $\left< c_{i \alpha}\right>= {\mu \over M}$,  $\left<c_{i \alpha} c_{j \beta}\right>= {\sigma_c^2 \over M} \delta_{ij} \delta_{\alpha \beta} + {\mu^2 \over M^2}\approx {\sigma_c^2 \over M} \delta_{ij} \delta_{\alpha \beta}$. 
	\item $K_\alpha= K+\delta K_\alpha $ with $\left< K_\alpha \right>=\frac{1}{M}\sum_\beta K_\beta=K$, $\left<\delta K_\alpha \delta K_\beta\right>=\delta_{\alpha \beta} \sigma_K^2.$   
	\item $m_i= m+\delta m_i $ with $\left<m_i\right>=m$, $\left< \delta m_i \delta m_j\right>=\delta_{ij}\sigma_m^2.$
	\item $\gamma=\frac{M}{S}$ and for the identity matrix $\gamma=1$.
\end{itemize}

Following similar steps as in \cite{advani2018statistical}, we perturb the ecosystem with a new species and resource $N_0$ and $R_0$. Ignoring $\mathcal{O}(1/M)$ terms yields the following equations: 
\begin{eqnarray} 
{d N_i \over dt  }\!&=& \!N_i\left[R_i\!-\!m\!+\! \sum_{\beta}(\frac{\mu}{M} + \sigma_c d_{i \beta}) R_{\beta}+(\frac{\mu}{M}\!+\!\sigma_c d_{i0})R_0\!-\!\delta m_i\right]\\
{d R_{\alpha} \over dt }\!&=&R_\alpha\left[\!K \!+\! \delta K_\alpha \!-\! R_\alpha -N_\alpha \!- \!\sum_j (\frac{\mu}{M}\!+\!\sigma_c d_{j\alpha})N_j-(\frac{\mu}{M}\!+\!\sigma_c d_{0\alpha})N_0\right]\\
{d N_0 \over dt} \!&=& \!N_0\left[R_0-\!m\!+\! \sum_{\beta}(\frac{\mu}{M}\!+\!\sigma_c d_{j\alpha}) R_{\beta}\!-\!\delta m_0\right]\\
{d R_0 \over dt} \!&=&\!R_0\left[K+ \delta K_0-R_0-N_0-\sum_{j}(\frac{\mu}{S}+\sigma_c d_{j 0}) N_j\right] \end{eqnarray} 

Denote by $\bar{N}_{\alpha /0}$, $\bar{R}_{\alpha /0}$ and $\bar{N}_i$, $\bar{R}_\alpha$ the equilibrium values of the species and resources
 before and after adding the newcomers, respectively. These can be related to each other using the susceptibilities defined above:
\begin{eqnarray}
\bar{N}_i &=& \bar{N}_{i /0} -\sigma_c\sum_{j}\nu^{N}_{ij} d_{j0}  R_0-\sigma_c\sum_{\beta}\chi^{N}_{i\beta} d_{0\beta}  N_0\\
\bar{R}_\alpha &=&\bar{R}_{\alpha /0} -\sigma_c\sum_i \nu^R_{\alpha i} d_{i 0}  R_0-\sigma_c\sum_\beta\chi^R_{\alpha\beta} d_{0\beta}  N_0
\end{eqnarray}

In what follows we assume Replica Symmetry. In this case, the sums in the equations above can be approximated as Gaussian random variables. For this reason, it is helpful to introduce new auxiliary random variables:
\begin{eqnarray} 
z_N&=&\sum_{\beta}\sigma_c \bar{R}_{\beta/0}d_{0\beta}-\delta m_0 \\
z_R&=&\sum_{j}\sigma_c \bar{N}_{j/0}d_{j0}-\delta K_0
\end{eqnarray}
where $ \left<z_N\right>=0$, $ \sigma_{z_N}=\sqrt{\sigma_c^2 q_R+\sigma_m^2}$ and  $\left<z_R\right>=0$, $ \sigma_{z_R}=\sqrt{\sigma_c^2 q_N+\sigma_K^2}$.

\textbf{Case 1:}  both $R_0$ and $N_0$ are positive. Following calculations analogous to \cite{advani2018statistical}  and noting that $\gamma=\frac{M}{S}=1$ yields:
\begin{equation}
\bar{R}_0=\text{max}\left[0, \frac{\sigma_c^2\chi(K-\mu \left< N\right>+z_R)-\mu \left< R\right>+m-z_N}{(1-\sigma_c^2\nu)\sigma_c^2\chi+1}\right]
\end{equation} 
\begin{equation}
\bar{N}_0=\text{max}\left[0, \frac{(1-\sigma_c^2\nu)(\mu \left< R\right>-m+z_N)+K-\mu \left< N\right>+z_R}{(1-\sigma_c^2\nu)\sigma_c^2\chi+1}\right]
\end{equation} 

\textbf{Case 2:}  either $R_0$ or $N_0$ is zero.  We get exactly the same expression as the random ecosystem we derived in  \cite{advani2018statistical}. 
\begin{equation}
\bar{R}_0=0, \quad   \bar{N}_0=\frac{\mu \left< R\right>-m+z_N}{\sigma_c^2\chi} \quad\text{or,} \quad \bar{N}_0=0, \quad \bar{R}_0= \frac{K-\mu \left< N\right>+z_R}{1-\sigma_c^2\nu}
\end{equation}

\textbf{Case 3:}  both $R_0$ and $N_0$ are zero, namely, 
\begin{equation}
\bar{R}_0=0 \text{ and }  \bar{N}_0=0. 
\end{equation} 
Combining the cases above, the steady state solution  is a Gaussian mixture depending on the positivity of $R_0$ and $N_0$.
\begin{equation}
\bar{R}_0=\Theta(R_0)\left[ \Theta(N_0) \frac{\sigma_c^2\chi(K-\mu \left< N\right>+z_R)-\mu \left< R\right>+m-z_N}{(1-\sigma_c^2\nu)\sigma_c^2\chi+1} + (1-\Theta(N_0)) \frac{K-\mu \left< N\right>+z_R}{1-\sigma_c^2\nu} \right]
\end{equation} 
\begin{equation}
\bar{N}_0=\Theta(N_0)\left[ \Theta(R_0)  \frac{(1-\sigma_c^2\nu)(\mu \left< R\right>-m+z_N)+K-\mu \left< N\right>+z_R}{(1-\sigma_c^2\nu)\sigma_c^2\chi+1} + (1-\Theta(R_0)) \frac{\mu \left< R\right>-m+z_N}{\sigma_c^2\chi} \right]
\end{equation} 

Cavity equations for the susceptibilities can be obtained directly by differentiating these equations:
\begin{eqnarray} 
\nu&=&\frac{1}{M}\sum_{i} \nu^{N}_{ii}=\left<\frac{\partial \bar{N}_0 }{\partial m} \right>=-\frac{\phi_N\phi_R(1-\sigma_c^2\nu)}{(1-\sigma_c^2\nu)\sigma_c^2\chi+1}-\frac{\phi_N(1-\phi_R)}{\sigma_c^2\chi}\label{nu}\\
\chi&=&\frac{1}{M}\sum_{\alpha} \chi^{R}_{\alpha\alpha}= \left<\frac{\partial\bar{R}_0 }{\partial K}\right>=\frac{\phi_N\phi_R \sigma_c^2\chi}{(1-\sigma_c^2\nu)\sigma_c^2\chi+1}+\frac{(1-\phi_N)\phi_R }{1-\sigma_c^2\nu}\label{chi}
\end{eqnarray}

\subsection{With resource extinction}\label{subsec:withrd}
Two solutions are found by solving  eq. (\ref{nu}) and eq. (\ref{chi}):
\begin{eqnarray}
\phi_R-\phi_N=0, \quad \chi=0, \quad \nu=\frac{1}{\sigma_c^2- 1}\label{CS1}
\end{eqnarray}
\begin{eqnarray}\label{CS2}
\phi_R-\phi_N> 0, \quad \chi=\phi_R-\phi_N, \quad \nu=\frac{1-2\phi_N\sigma_c^2 +\phi_R\sigma_c^2-\sqrt{1+2(1-2\phi_N)\phi_R\sigma_c^2+\phi_R^2\sigma_c^4}}{2\sigma_c^4(\phi_R-\phi_N)}.
\end{eqnarray}
 
\subsection{Without resource extinction}\label{subsec:withoutrd}
In this case, the resource never vanishes so that we can fix $\phi_R=1$ and solve eq. (\ref{nu}) and eq. (\ref{chi}). Two solutions are found:
\begin{eqnarray}\label{CS3}
1-\phi_N=0, \quad \chi=0, \quad \nu=\frac{1}{\sigma_c^2- 1}
\end{eqnarray}
\begin{eqnarray}\label{CS4}
1-\phi_N> 0, \quad \chi=1-\phi_N, \quad \nu=\frac{1-2\phi_N \sigma_c^2+\sigma_c^2-\sqrt{1+2\sigma_c^2-4\phi_N\sigma_c^2+\sigma_c^4}}{2\sigma_c^4(-1+\phi_N)}.
\end{eqnarray}
Above two solutions are continuous at the transition point: $\chi=0$ i.e. $\phi_N=1$.  Assume there is a small perturbation near the transition: $\phi_N=1-\epsilon$ and $\epsilon\ll1$ and $\nu$ in eq. (\ref{CS4}) can be expanded around $\epsilon$. It is easy to check the $\nu$ in eq. (\ref{CS4})  has the same expression as eq. (\ref{CS3}) at the first order of $\epsilon$. 
Therefore, only one solution exists:
\begin{eqnarray}\label{CS5}
 \chi=1-\phi_N, \quad \nu=\frac{1-2\phi_N \sigma_c^2+\sigma_c^2-\sqrt{1+2\sigma_c^2-4\phi_N\sigma_c^2+\sigma_c^4}}{2\sigma_c^4(-1+\phi_N)}
\end{eqnarray}
The comparison between cavity solutions and numerical simulations for $\chi$ and $\nu$ are given in Figure \ref{chi_si} and  Figure \ref{rdelete} respectively. 

\subsection{Without resource extinction and species extinction}
In this case, both the resource and the species never vanish so that we can fix $\phi_R=1$ and $\phi_N=1$. Solving eq. (\ref{nu}) and eq. (\ref{chi}), only one solution is found:
\begin{eqnarray}\label{CS6}
 \chi=0, \quad \nu=\frac{1}{\sigma_c^2- 1}.
\end{eqnarray}

\subsection{Behavior in Three Regimes}
To understand these solutions and behaviors better, it is helpful to consider three regimes: \textit{Regime A} where $\chi=\phi_R-\phi_N=0$, \textit{Regime B} where $\chi$ becomes nonzero  and species start to extinct, and
\textit{Regime C} where $\sigma_c \gg 1$ and it becomes a random ecosystem. 

In \textit{Regime B}, resource extinction has a significant effect on the system's feasibility, shown in Figure \ref{rdelete}.
With resource extinction, equation (\ref{CS2}) shows there is a sudden change for the linear response function  $\nu$ from \textit{Regime A}: $\chi=0$  to \textit{Regime B} $\chi\neq0$.  As $\nu\sim \frac{1}{\phi_R-\phi_N}$, even a slightly decrease of the number of surviving species will induce a huge perturbation to the ecosystem, corresponding to a phase transition between \textit{Regime A} and \textit{Regime B} at $\sigma_c^*\sim0.2$.

Without resource extinction, equation (\ref{CS5}) shows the linear response function  $\nu$ is continuous from \textit{Regime A} to \textit{Regime B}. There is a crossover instead of a phase transition there. The peak for the crossover is a finite value and can be calculated by taking the derivative of  equation (\ref{CS5})  over $\sigma_c$, ignoring the correlation between $\sigma_c$ and $\phi_N$. It happens approximately at $\sigma_c^*= \sqrt{4\phi_N-2}\sim 1.04$, where $\phi_N=0.77$ can be obtained from numerical simulation.  The explanation for the difference from random matrix theory are provided in the main text and also the spectrums in Figure \ref{pdf} and Figure \ref{rdelete}. 

Without resource and species extinction, as shown in equation (\ref{CS6}), $\nu$ diverges at $\sigma_c^*=1$, corresponding to $\lambda_{\rm min}$ reaching exactly zero. This result is also consistent with equation (\ref{st}), predicted by random matrix theory, which ignores the effect of row or column deletions in the interaction matrix. This tells there do not  exists any feasible solutions for the coexistence of M species and M resources.  Therefore species must go extinct before $\sigma_c^*=1$. 

In \textit{Regime C}, further increasing of $\sigma_c$ after $\sigma_c>1$,  the $\sigma_c^4$ term in the square root becomes dominating and the the susceptibility $\nu$ behaves like a random ecosystem quickly, which explains the dramatic drop of the species packing shown in Figure \ref{heatmap}.  It indicates the ecosystem tends to a self-organized random state. 

\subsection{Solutions in Regime A and C}
In \textit{Regime A}($\sigma_c\ll1$), 
for eqs. (\ref{Ma}) with resource extinction, the solutions for the steady-states become,
\begin{eqnarray}
R_0=\text{max}\left[0, m-z_N\right], \quad N_0=\text{max}\left[0,K+z_R\right].
\end{eqnarray}
For eqs. (\ref{Ma1}) without resource extinction, the solutions for the steady-states become,
\begin{eqnarray}
R_0=m-z_N, \quad N_0=\text{max}\left[0,K+z_R\right].
\end{eqnarray}
For ecosystems without resource and species extinction, the solutions for the steady-states become,
\begin{eqnarray}
R_0=m-z_N, \quad N_0=K+z_R.
\end{eqnarray}

For \textit{Regime C}  ($\sigma_c\gg1$), 
for eqs. (\ref{Ma}) with resource extinction, the solutions for the steady-states become,
\begin{eqnarray}
R_0=\text{max}\left[0, \frac{K-\mu \left< N\right>+z_R}{1-\sigma_c^2\nu}\right], \quad
N_0=\text{max}\left[0, \frac{\mu \left< R\right>-m+z_N}{\sigma_c^2\chi}\right]\label{Nocavity},
\end{eqnarray}
in agreement with the equations obtained in \cite{advani2018statistical} for purely random interactions. For equations. (\ref{Ma1}) without resource extinction, the solutions for the steady-states become,
\begin{eqnarray}
R_0= \frac{K-\mu \left< N\right>+z_R}{1-\sigma_c^2\nu},\quad
N_0=\text{max}\left[0, \frac{\mu \left< R\right>-m+z_N}{\sigma_c^2\chi}\right]\label{Nocavity1}.
\end{eqnarray}
For ecosystems without resource and species extinction, the solutions for the steady-states become,
\begin{eqnarray}
R_0= \frac{K-\mu \left< N\right>+z_R}{1-\sigma_c^2\nu}, \quad
N_0=\frac{\mu \left< R\right>-m+z_N}{\sigma_c^2\chi}\label{Nocavity2}.
\end{eqnarray}

\section{Lotka-Volterra Model, Wishart Matrix and Marchenko-Pastur Law }\label{sec:mapLV}
In this section, we show how the generalized Lotka-Volterra model can be related to the CRM, and in particular, the how the steady states of the
two models can be made to coincide. Solving for the steady-state values
of the non-extinct resources by setting the bottom equation in (\ref{Ma}) equal to zero gives:
$$
\bar{R}_\alpha=K_\alpha-\sum_i N_i \bar{C}_{i\alpha}
$$
Substituting this into the top equation in (\ref{Ma}) gives:
$$
\frac{dN_i}{dt}=N_i\left (\sum_{\alpha\in \mathbf{M^*}} C_{i \alpha}K_\alpha-m_i- \sum_j A_{ij}N_j \right)
\label{effLV}
$$
where we have defined an interaction matrix $A_{ij}=\sum_{\alpha\in \mathbf{M^*}}\bar{C}_{i\alpha}\bar{C}^{T}_{\alpha j}$ and $\mathbf{M^*}$  is the set of surviving resources.
We can use this equation to solve for the steady-state (equilibrium) abundances of non-extinct species, and arrive at the expression:
$$
\bar{N}_i= \sum_{j\in \mathbf{S}^*} A^{-1}_{ij}(\sum_{\alpha\in \mathbf{M^*}} C_{j \alpha}K_\alpha-m_j)
$$
where $\mathbf{S}^*$ is the set of surviving species. In terms of $\bar{N}_i$, the Lotka-Volterra equations become:
\begin{eqnarray}\label{LVsi}
\frac{d N_i}{dt}&=&-\bar{N}_i \sum_j A_{ij}(N_j-\bar{N}_j)
\end{eqnarray}
with community matrix 
\begin{align}
J_{ij} = \left(\frac{\partial}{\partial N_j} \frac{d N_i}{dt}\right)_{\{\bar{N}_j\}} = -\bar{N}_i A_{ij}.
\end{align}

In May's work, $J_{ij}$ is assumed to be an i.i.d. random matrix and an extension of Wigner's arguments about Gaussian random matrices is used to compute the leading eigenvalue \cite{may1972will}. Since the $\bar{N}_i$ are not known \emph{a priori}, the stability of Lotka-Volterra type dynamics are more easily studied in terms of the eigenvalues of $A_{ij}$, using the connection between the leading eigenvalues of $\mathbf{J}$ and $\mathbf{A}$ derived below.

\section{Relating the eigenvalues of $\mathbf{A}$ and $\mathbf{J}$}\label{sec:Jacobimatrix}
In this section, we prove that the largest eigenvalue $\lambda_{\rm max}$ of the community matrix $\mathbf{J}$ (which controls the Lyapunov stability of the fixed point) is negative if and only if the smallest eigenvalue $\lambda_{\rm min}$ of the Lotka-Volterra competition matrix $\mathbf{A}$ is positive. For this stability analysis, we remove the rows and columns corresponding to species that go extinct in the steady state, since allowing $N_i = 0$ trivially generates zero eigenvalues. $\mathbf{J}$ and $\mathbf{A}$ will always refer to the resulting matrices of dimension $S^* \times S^*$. 

We start by defining the diagonal matrix $\mathbf{\bar{N}}$, whose nonzero elements are the equilibrium population sizes $\bar{N}_i$. This lets us write
\begin{eqnarray}
\mathbf{J}=- \mathbf{\bar{N}}^{1/2} (\mathbf{\bar{N}}^{1/2} \mathbf{A}\mathbf{\bar{N}}^{1/2})\mathbf{\bar{N}}^{-1/2}
\end{eqnarray}
where $\mathbf{\bar{N}}^{1/2}$ is the diagonal matrix whose entries are the square roots of the population sizes. This equation says that $\mathbf{J}$ is similar to $-\mathbf{W} \equiv -\mathbf{\bar{N}}^{1/2} \mathbf{A}\mathbf{\bar{N}}^{1/2}$, which implies that they share the same eigenvalues. 

Since $\mathbf{W}$ and $\mathbf{A}$ are both symmetric matrices, their eigenvalues are all real, and the positivity of all the eigenvalues is equivalent to the positive-definiteness of the matrix. 

Now we note that $\mathbf{W}$ is positive definite if and only if $\mathbf{A}$ is positive definite. For if $\mathbf{A}$ is positive definite, then $\mathbf{x}^T \mathbf{A} \mathbf{x}>0$ for all column vectors $\mathbf{x}\neq 0$, including the column vector $\mathbf{x} = \mathbf{\bar{N}}^{1/2}\mathbf{y}$ for any column vector $\mathbf{y}\neq 0$. But this implies that $\mathbf{y}^T \mathbf{\bar{N}}^{1/2} \mathbf{A}\mathbf{\bar{N}}^{1/2} \mathbf{y}>0$ for all $\mathbf{y} \neq 0$, i.e., that $\mathbf{W}$ is positive definite. Conversely, if $\mathbf{W}$ is positive definite, then $\mathbf{y}^T \mathbf{\bar{N}}^{1/2} \mathbf{A}\mathbf{\bar{N}}^{1/2} \mathbf{y}>0$ for all $\mathbf{y} \neq 0$, including $\mathbf{y} = \mathbf{\bar{N}}^{-1/2} \mathbf{x}$ for any $\mathbf{x}\neq 0$. But this implies that $\mathbf{x}^T \mathbf{A}\mathbf{x} >0$ for all $\mathbf{x}\neq 0$, i.e., that $\mathbf{A}$ is positive definite.

We conclude that the eigenvalues of $\mathbf{W}$ are all positive if and only if the eigenvalues of $\mathbf{A}$ are all positive. Therefore the largest eigenvalue of $\mathbf{J} = -\mathbf{\bar{N}}^{1/2}\mathbf{W}\mathbf{\bar{N}}^{-1/2}$ is negative if and only if the smallest eigenvalue of $\mathbf{A}$ is positive, as claimed in the main text.

An alternative but much simpler proof can be provided with the properties of the D-stable matrix\cite{hogben2013handbook}. A real square matrix $\mathbf{A}$ is said to be D-stable if the matrix $\mathbf{DA}$ is positive definite for every choice of a positive diagnoal matrix $\mathbf{D}$. A sufficient condition for D-stablility is that $\mathbf{A}+\mathbf{A^T}$ is positive definite. The Lotka-Volterra competition matrix $\mathbf{A}$ is symmetric and positive definite, $i.e.$, D-stable. $\mathbf{\bar{N}}$ is all positive. It is obvious that $\mathbf{J}=\mathbf{\bar{N}A}$ is positive definite. Further discussions about its application in ecology can be found in \cite{grilli2017feasibility, servan2018coexistence}

\section{Correspondence between RMT and cavity solution}\label{RMT}
Our numerical simulations show that after the transition, our ecosystems are well described by purely random interactions. This suggests
that we should be able to derive our cavity results using Random Matrix Theory (RMT). We now show that this is indeed the case. Our starting 
point are the average susceptibilities which are defined as:
\begin{eqnarray}
\chi&=&\frac{1}{M}\sum_{\alpha\in \mathbf{M}}\chi^R_{\alpha \alpha}=\frac{1}{M}\sum_{\alpha\in \mathbf{M^*}}\chi^R_{\alpha \alpha} \\
\nu&=&\frac{1}{S}\sum_{i \in \mathbf{S}}\nu^N_{i i}=\frac{1}{S}\sum_{i \in \mathbf{S^*}}\nu^N_{i i}.
\end{eqnarray}
From the cavity calculations, we only care about $\chi^R_{\alpha \beta}$ and $\nu^N_{i j}$, because the other susceptibilities are lower order in $1/M$. 

We can combine these equations with (\ref{chi1}) and (\ref{nu1}) to obtain
\begin{eqnarray}
\chi&=&\frac{1}{M}\sum_{\alpha\in \mathbf{M^*}}\chi^R_{\alpha \alpha}
=\frac{1}{M}\text{Tr}(\chi^R_{\alpha \beta})\\
&=&\frac{1}{M}\text{Tr}(\delta_{\alpha\beta})-\frac{1}{M}\text{Tr}\left(\sum_{i\in \mathbf{S^*}}\sum_{j\in \mathbf{S^*}} \bar{C}^T_{\alpha i}A^{-1}_{ij}\bar{C}_{j\beta}\right)\nonumber\\
&=&\frac{M^*}{M}-\frac{1}{M}\text{Tr}\left(\sum_{i\in \mathbf{S^*}}\sum_{j\in \mathbf{S^*}} A^{-1}_{ij}\bar{C}_{j\beta}\bar{C}^T_{\beta h}\right)\nonumber\\
&=& \frac{M^*}{M}-\frac{S^*}{M}=\phi_R-\gamma^{-1}\phi_N\label{chiRMT}
\end{eqnarray}

We now show that the cavity solutions are consistent with results from RMT using equations  (\ref{chi1}) and(\ref{nu1}) in Regime A and Regime C described in the main text.

\subsection{Regime A: $\bar{\mathbf{C}}=\mathds{1}$}
This regime happens when $\sigma_c\ll 1$. Substituting,  $\bar{\mathbf{C}}=\mathds{1}$ into equations (\ref{chi1}) and (\ref{nu1}) yields
\begin{equation}
\chi=0, \quad \nu=-1.
\end{equation}
This is consistent with the cavity solution equation (\ref{CS1}) with $\sigma_c=0$ since in this case $S^*=S=M$.

\subsection{Regime C: $\bar{C}_{i\alpha} \textit{ i.i.d. } \mathcal{N}(0, \sigma_c/\sqrt{M})$}
In this regime, $\sigma_c\gg 1$. In this case, $A_{ij}=\sum_{\alpha\in \mathbf{S^*}} \bar{C}_{i\alpha}\bar{C}^T_{\alpha j}$ takes the form of a Wishart Matrix. We
will exploit this to calculate $\chi$ and $\nu$. Notice,
\begin{eqnarray}
\nu&=&\frac{1}{S}\sum_{i \in \mathbf{S^*}}\nu^N_{i i}=-\frac{1}{S}\text{Tr}(A^{-1}_{ij})=-\frac{1}{S}\sum^{S^*}_{i=1} \lambda^{-1}_i
\end{eqnarray}
where $\lambda_i$ is the eigenvalue of $A_{ij}$.
From the Marchenko-Pastur law \cite{marchenko1967distribution}, we know that the eigenvalues of a random Wishart matrix obey the
Marchenko-Pastur distribution. Substituting equation (\ref{MPlawsi}) into the expression for $\nu$ and
replacing the sum with an integral yields:
\begin{eqnarray}
\nu&=&-\frac{S^*}{S}\int_a^b\frac{1}{x} \rho(x)dx\label{intMP}
\\&=&-\frac{S^*}{S}\frac{a+b-2\sqrt{ab}}{4\sigma_c^2y \sqrt{ab}}\nonumber\\
&=&-\frac{1}{\sigma_c^2}\frac{\phi_N}{\phi_R-\gamma^{-1}\phi_N}\nonumber
\end{eqnarray}
The second line of equation (\ref{intMP}) is obtained by transferring the integral function to a complex analytic function and applying the residue theorem. This result is the same as the cavity solution equation (\ref{CS2}) when $\sigma_c\gg1$.

\subsection{Regime B using the Stieltjes transformation}\label{ipnm}
In Regime B, it hard to estimate the minimum eigenvalue.  We can use Stieltjes transformation of information-plus-noise-type matrices which are well studied in wireless communications\cite{dozier2007empirical, couillet2011random,loubaton2011almost}, where $
\mathbf{B}$ represents the information encoded in the signal and $\mathbf{C}$ is the noise in wireless communications.  In this case, we have
$$\bar{C}_{i \alpha} = \mathds{1} + C_{i\alpha}, \quad C_{i\alpha}  \textit{ i.i.d. } \mathcal{N}(0, \sigma_c/\sqrt{M}).$$ 
\begin{eqnarray}\label{noisemodel}
A_{ij}=\sum_{\alpha\in M^*}\bar{C}_{i\alpha}\bar{C}^{T}_{\alpha j}
=\sum_{\alpha\in M^*}C_{i\alpha}C^{T}_{\alpha j}+ C_{i\alpha}+C^T_{\alpha i}+\mathds{1}
\end{eqnarray}

\begin{figure}
	\centering
	\includegraphics[width=0.6\textwidth]{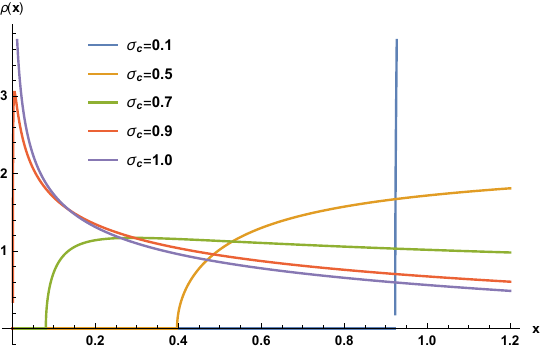}
	\caption{ The asymptotic spectrum of $A_{ij}$ for different values of $\sigma_c$ by solving equation (\ref{st1}) numerically.}
	\label{mz}
\end{figure}

Using \textbf{Theorem 1.1} in \citeauthor{dozier2007empirical}\cite{dozier2007empirical}, the Stieltjes transform  $m(z)$ of $A_{ij}$ satisfies
\begin{eqnarray}\label{st}
\sigma_c^4 z m^3-2\sigma_c^2 z m+(\sigma_c^2+z-1)m -1=0
\end{eqnarray}
The asymptotic spectrum of $A_{ij}$ can be obtained by $m(z)$, the solution of  equation (\ref{st}) with
\begin{eqnarray}\label{st1}
\rho(x)=\lim_{\varepsilon\to 0^+}\frac{m(x-i\varepsilon)-m(x+i\varepsilon)}{2i\pi}
\end{eqnarray}
The result is shown in Figure \ref{mz}.  The minimum eigenvalue reaches 0 nearly at $\sigma^*_c=1$, as predicted by the cavity solution. 

We emphasize that the phase transition point, derived from eq. (\ref{st}), does not change at different $\mu$. In the original paper by \citeauthor{marchenko1967distribution}, eq. (\ref{MPlaw}) requires the elements are i.i.d variables with mean 0 and variance $\sigma^2$.  Recently, it has been shown a nonzero $\mu$ contributes only one eigenvalue $\lambda$. It is either in the domain of MP Law,  $\lambda \in  [a, b]$ or off the domain  $\lambda >b$\cite{baik2005phase, benaych2012singular} and thus it does not affect the minimum eigenvalue.
We can understand it intuitively with a simple example: $\mathds{1}+\mu \mathds{J}$,  where $\mathds{1}$ is the identity matrix, $\mathds{J}$ is a $n\times n$ all-ones matrix. The eigenvalues can be calculated by
$$\mathrm{Det}((1-\lambda)\mathds{1}+\mu \mathds{J})=[1-\lambda+(n-1)\mu](1-\lambda)^{n-1}.$$
It shows when $n\gg1$, it only contributes a very large eigenvalue $1+(n-1)\mu$ and the others stay at 1.


\section{Parameters in simulations}\label{SI:simulations}
\begin{figure}
\centering
\includegraphics[width=0.6\textwidth]{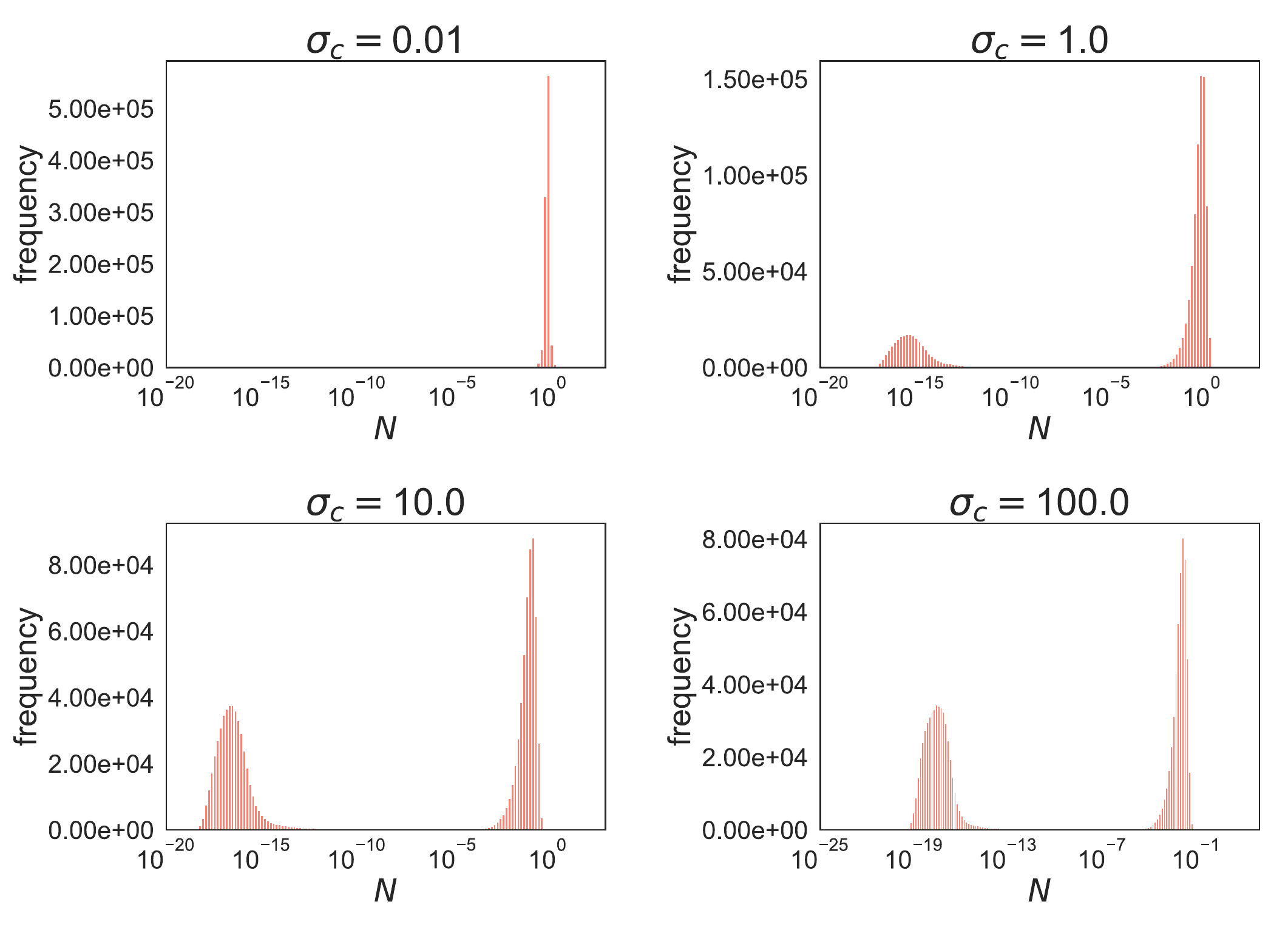}
\caption{Species abundance $N$ in equilibrium at different $\sigma_c$. The simulation details can be found at  Appendix \ref{SI:simulations}.}
\label{Nsim}
\end{figure}

\begin{figure}[h]
	\centering
	\includegraphics[width=0.5\textwidth]{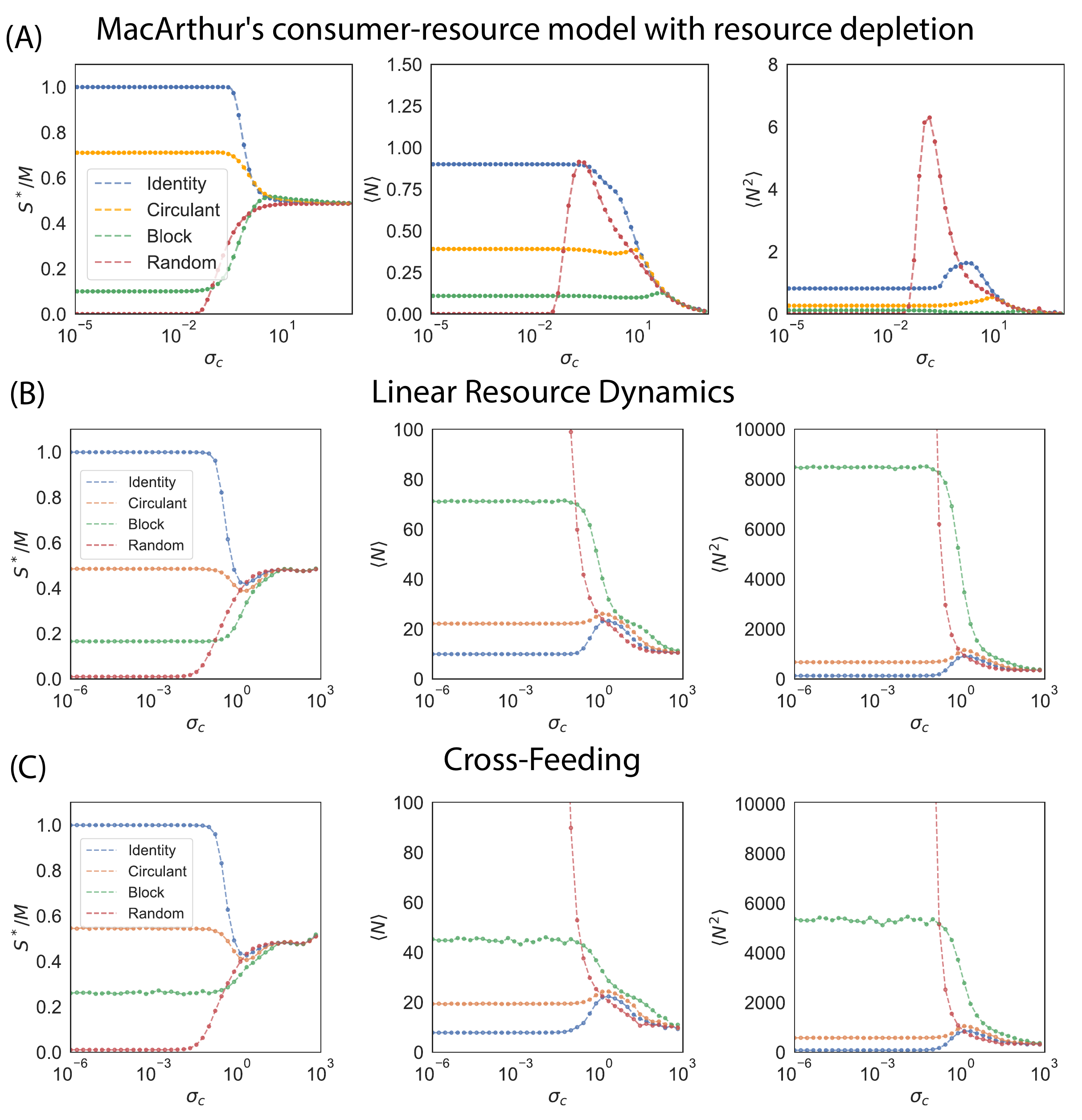}
	\caption{Community properties for generalized consumer-resource models under Gaussian noise. \textbf{(A)} MacArthur's consumer resource model with resource extinction. \textbf{(B)} Linear resource dynamics: the resource dynamics is changed to $\frac{dR_\alpha}{dt}=K_\alpha- R_\alpha-\sum_iN_iC_{i \alpha}R_\alpha$. \textbf{(C)} With cross-feeding: the dynamics is described in equation (17) in Supplementary Information of \cite{marsland2018available}. The noise is only applied on the consumption matrix and $D$ is kept the same at different $\sigma_c$. In both models, $\mathbf{B}=\mathds{1}$. See Appendix \ref{SI:simulations} for details.}
	\label{dynamics}
\end{figure}

\begin{figure}[h]
	\centering
	\includegraphics[width=0.5\textwidth]{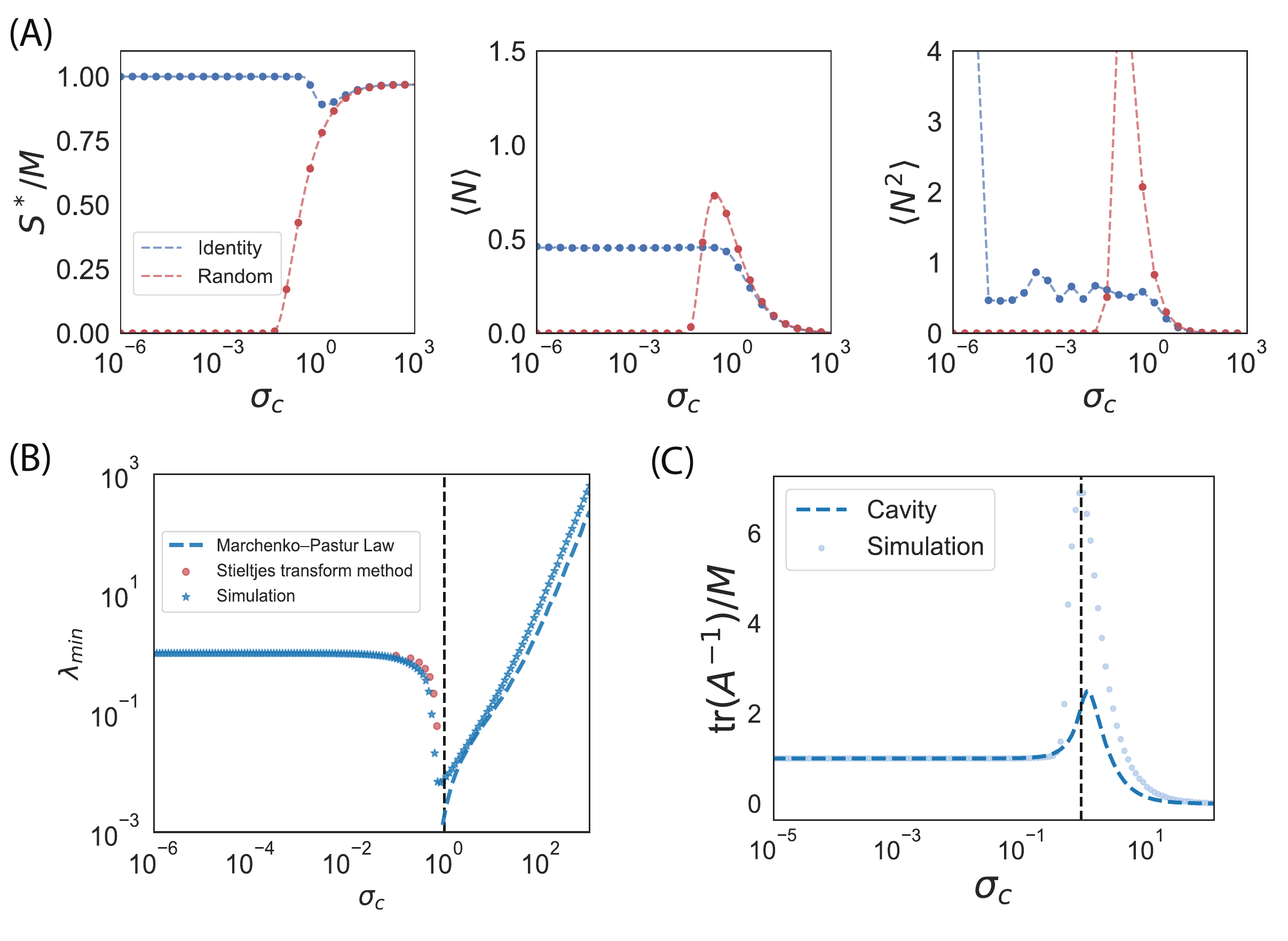}
	\caption{Effects when $S\neq M$. \textbf{(A)} Community properties \textbf{(B)} the minimum eigenvalue $\lambda_{min}$.  \textbf{(C)}  the mean sensitivity $\nu$. All simulations are the same as figures in the main text except $S=200, M=100$. See Appendix \ref{SI:simulations} for details.}
	\label{size}
\end{figure}

\begin{figure}[h]
	\centering
	\includegraphics[width=0.8\textwidth]{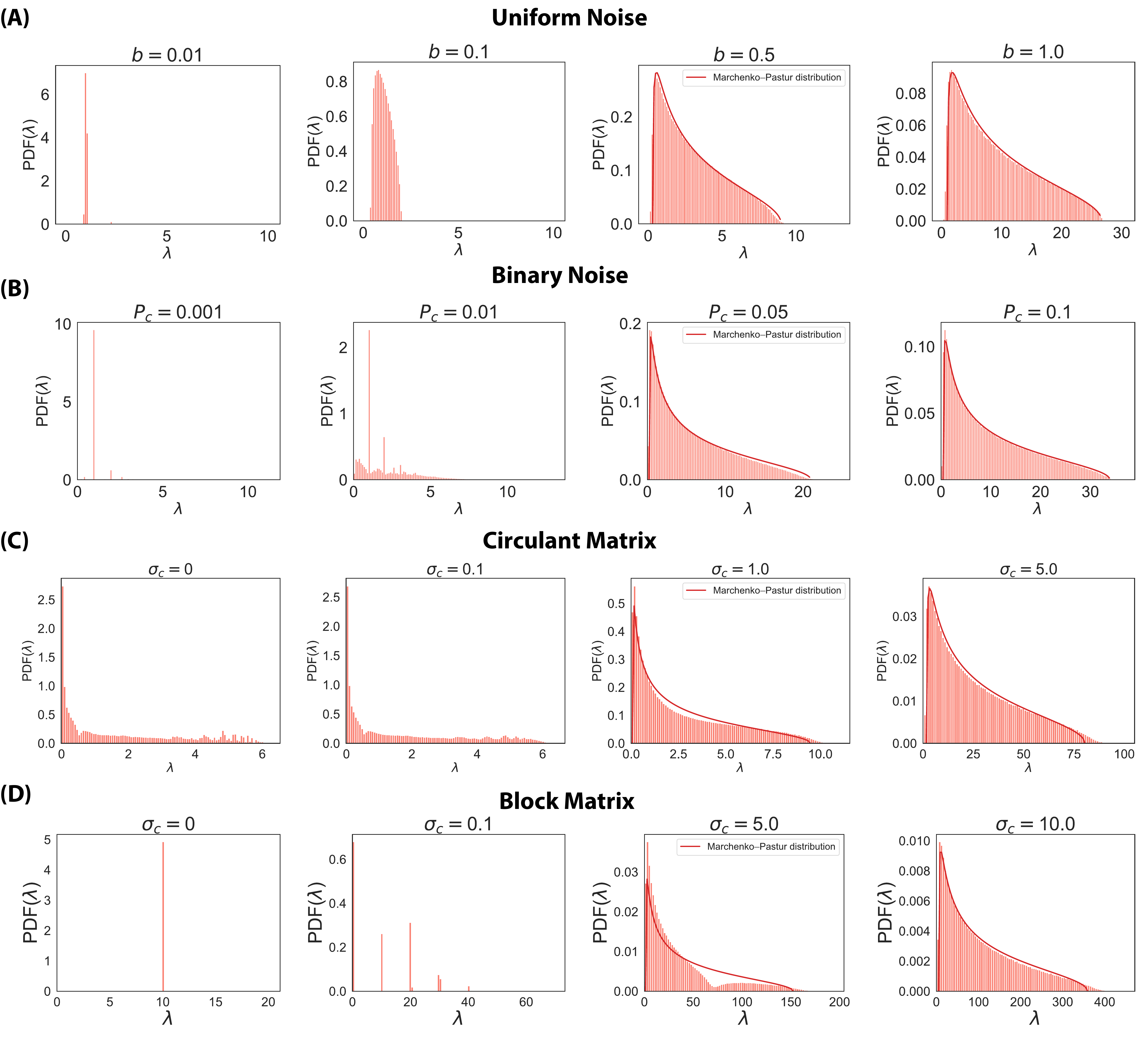}
	\caption{Spectra of $A_{ij}$ in different cases. \textbf{(A)} Uniform Noise: $\mathcal{U}(0,b)$ and \textbf{(B)} Binary Noise: ${Bernoulli}(p_c)$; the engineered matrix $\mathbf{B}$ is an identity matrix.  \textbf{(C)} Gaussian noise and the engineered matrix $\mathbf{B}$ is a circulant matrix. \textbf{(D)} Gaussian noise and the engineered matrix $\mathbf{B}$ is a block matrix. Note that $A_{ij}$ are obtained from numerical simulations. See Appendix \ref{SI:simulations} for details.}
	\label{matrix}
\end{figure}

\begin{figure}[h]
	\centering
	\includegraphics[width=0.7\textwidth]{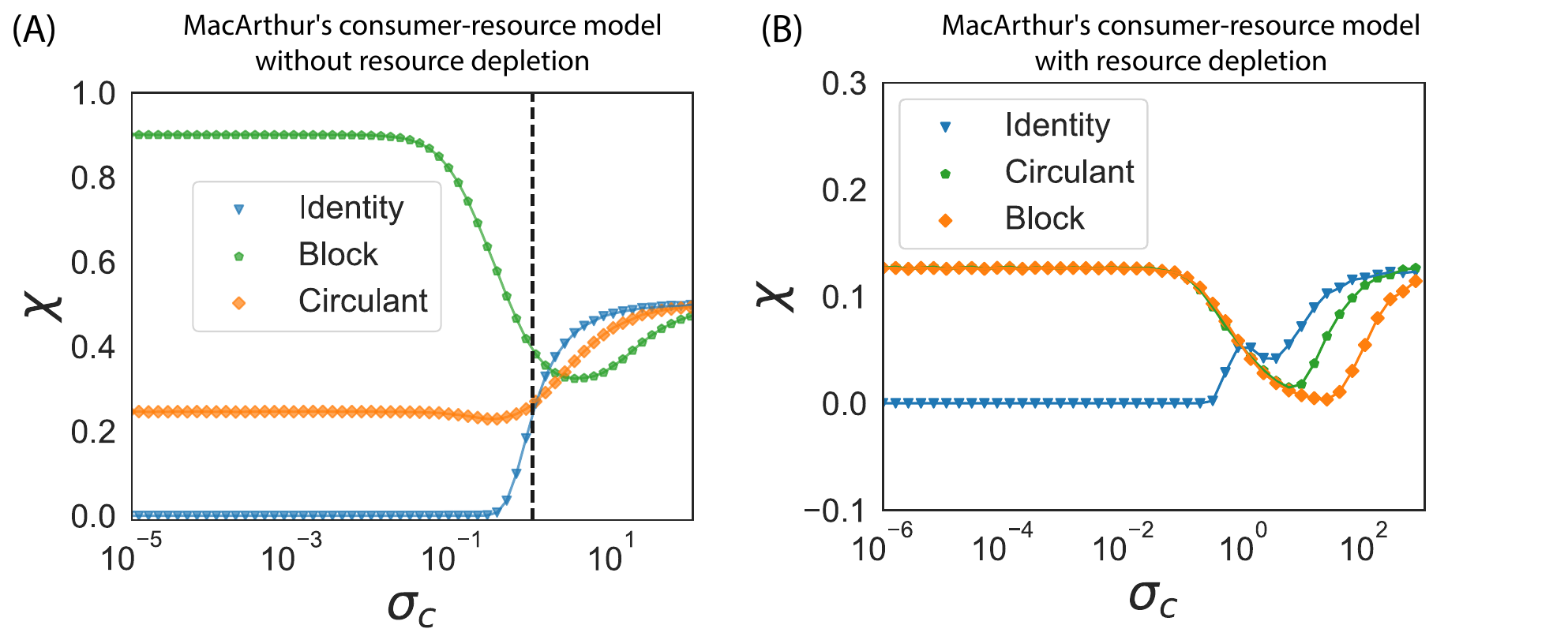}
	\caption{ Comparison between numerical simulations(scatter points) and cavity solutions(solid lines) for $\chi$ at different $\sigma_c$ for different cases.   \textbf{(A)} CRM without resource extinction, eqs. (\ref{Ma1}).  \textbf{(B)} CRM with resource extinction, eqs. (\ref{Ma}). Note $S^*$ and $M^*$ are obtained from the numerical simulations, although in principle they could be obtained by solving the cavity equations directly. }
	\label{chi_si}
\end{figure}

\begin{figure}[h]
	\centering
	\includegraphics[width=0.6\textwidth]{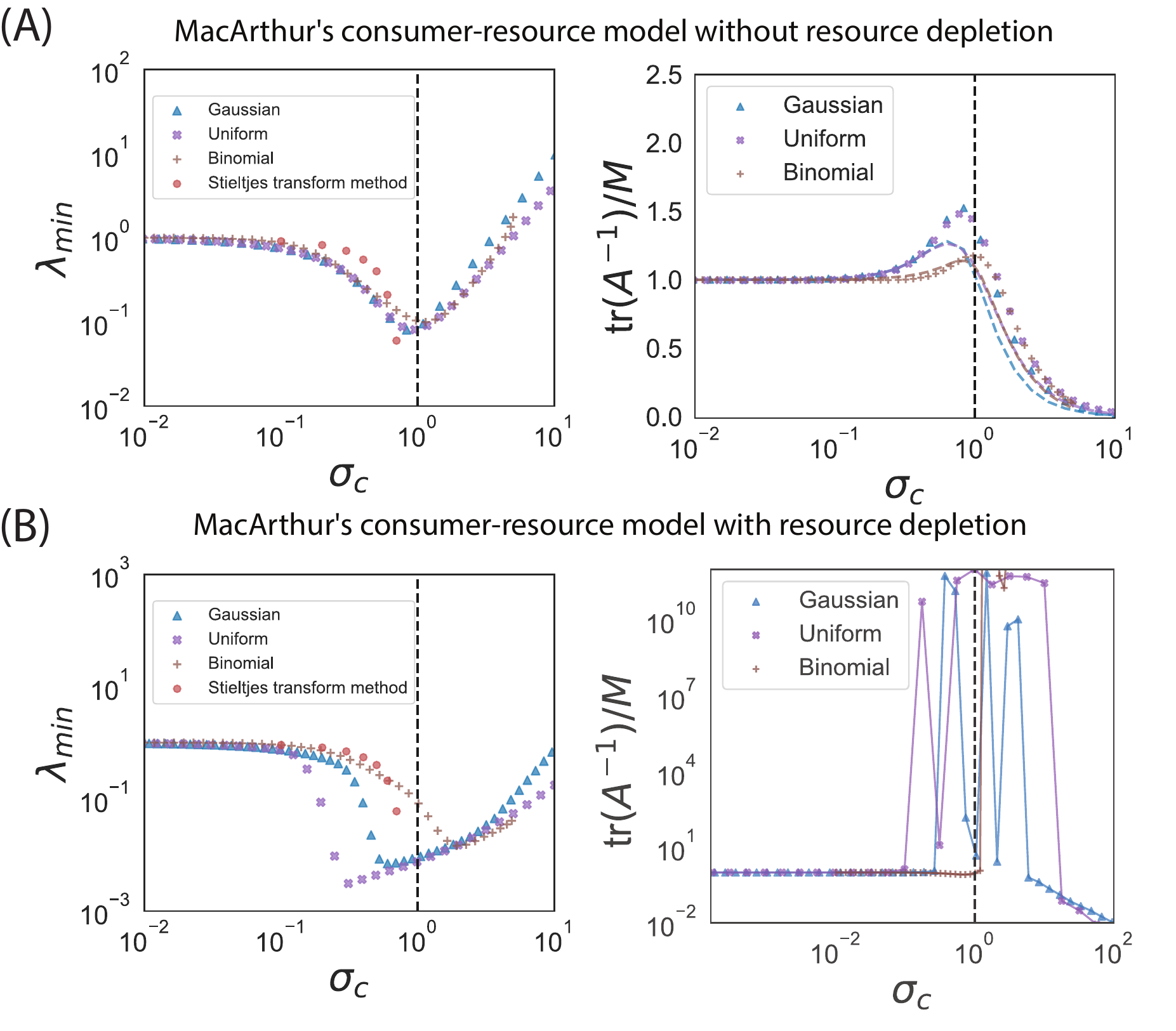}
	\caption{Comparison the minimum eigenvalue $\lambda_{min}$ and  the mean sensitivity $\nu$ between different distributions for the identity case at different $\sigma_c$.  \textbf{(A)} CRM without resource extinction, eqs. (\ref{Ma1}).  \textbf{(B)} CRM with resource extinction, eqs. (\ref{Ma}). Note that  the Bernoulli and uniform distribution to are mapped the corresponding Gaussian distribution $\mu=p_cM$, $\sigma_c=\sqrt{Mp_c(1-p_c)}$ and $\mu=bM/2$, $\sigma_c=b\sqrt{M/12}$, respectively. }
	\label{distributions}
\end{figure}

\begin{figure}[h]
	\centering
	\includegraphics[width=0.6\textwidth]{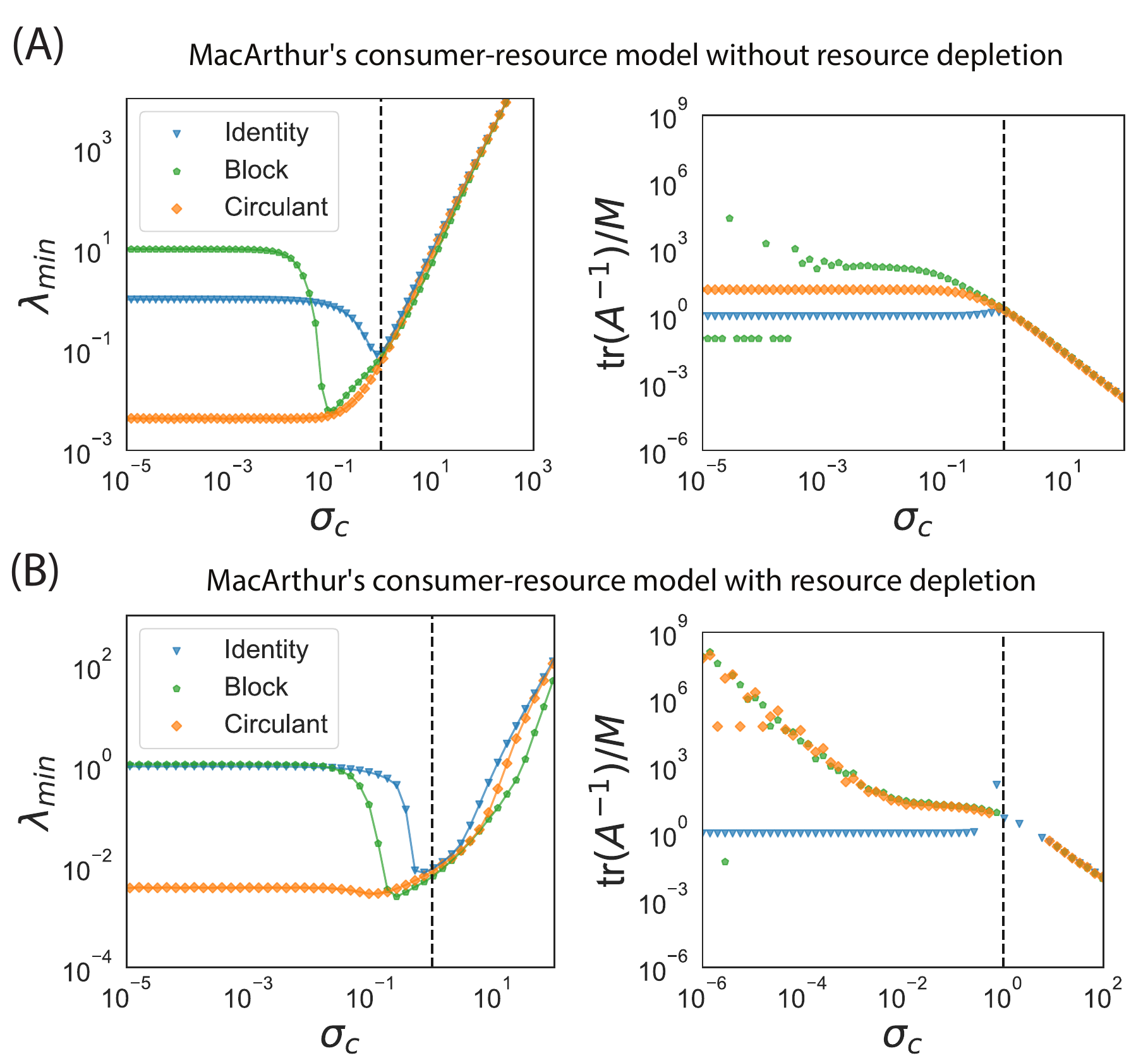}
	\caption{Comparison the minimum eigenvalue $\lambda_{min}$ and  the mean sensitivity $\nu$ between different engineered matrices $\mathbf{B}$ at different $\sigma_c$.   \textbf{(A)} CRM without resource extinction, eqs. (\ref{Ma1}).  \textbf{(B)} CRM with resource extinction, eqs. (\ref{Ma}). }
	\label{structures}
\end{figure}

All simulations are done with the CVXPY package\cite{cvxpy_rewriting}. The code is available on GitHub at \url{https://github.com/Emergent-Behaviors-in-Biology/typical-random-ecosystems}.
\begin{itemize}
\item Figure \ref{heatmap}(B), \ref{phase}(B), \ref{pdf}(C, D): the consumer matrix $\mathbf{C}$ is sampled from the Gaussian distribution $\mathcal{N}(\frac{\mu}{M},\frac{\sigma_c}{\sqrt{M}} )$. $S=100$, $M=100$, $\mu=0$, $K=1$, $\sigma_K=0.1$ ,  $m=0.1$, $\sigma_m=0.01$,  and each data point is averaged from 5000 independent realizations.  The model is simulated with eqs. (\ref{Ma1}).

\item Figure \ref{phase}(C): the consumer matrix $\mathbf{C}$ is sampled from the uniform distribution $\mathcal{U}(0, b)$.  $S=100$, $M=100$, $\mu=0$, $K=1$, $\sigma_K=0.1$ ,  $m=0.1$, $\sigma_m=0.01$,  and each data point is averaged from 5000 independent realizations.   The model is described by eqs. (\ref{Ma1}).

\item Figure \ref{phase}(D): the consumer matrix $\mathbf{C}$ is sampled from the Bernoulli distribution $\mathit{Bernoulli}(p_c)$.  $S=100$, $M=100$, $\mu=0$, $K=1$, $\sigma_K=0.1$ ,  $m=0.1$, $\sigma_m=0.01$,  and each data point is averaged from 5000 independent realizations.   The model is described by eqs. (\ref{Ma1}).

\item Figure \ref{pdf}(B), \ref{matrix}: the simulation is the same as Fig. \ref{phase}(B). Each spectrum is drawn from 10000 independent realizations. 

\item Figure \ref{rdelete}: the consumer matrix $\mathbf{C}$ is sampled from the Gaussian distribution $\mathcal{N}(\frac{\mu}{M},\frac{\sigma_c}{\sqrt{M}} )$. $S=100$, $M=100$, $\mu=0$, $K=1$, $\sigma_K=0.1$ ,  $m=0.1$, $\sigma_m=0.01$. The model without resource extinction simulated with eqs. (\ref{Ma1}),  and each data point is averaged from 5000 independent realizations.. The model with resource extinction is simulated with eqs. (\ref{Ma}),  and each data point is averaged from 4000 independent realizations. Each spectrum is drawn from 1 independent realizations for $S=500$.

\item Figure \ref{Nsim} : the simulation is the same as Fig. \ref{phase}(B). Each histogram is drawn from 10000 independent realizations. 

\item Figure \ref{dynamics}: the consumer matrix $\mathbf{C}$ is sampled from the Gaussian distribution $\mathcal{N}(\frac{\mu}{M},\frac{\sigma_c}{\sqrt{M}} )$. $S=100$, $M=100$, $\mu=1$, $K=1$, $\sigma_K=0.1$ ,  $m=0.1$, $\sigma_m=0.01$. For (C): $\omega=1$, $\sigma_\omega=0$, model details can be found in \cite{cui2019effect}; For (D), the dynamics is described in equation (17) in Supplementary Information of \cite{marsland2018available}. The noise is only applied on the consumption matrix and $D$ is kept the same at different $\sigma_c$. Each data point is averaged from 4000 independent realizations for (A), from 5000 independent realizations for (B) and 1000 independent realizations for (C, D). 

\item Figure  \ref{size}: the simulation is the same as Figure \ref{heatmap}(B) except $S=200$, $M=100$. For the identity case, the consumer matrix is obtained by concatenating the $M\times M$ identity plus noise matrix and a $(S-M)\times M$ gaussian random matrix. The model without resource extinction simulated with eqs. (\ref{Ma1}),  and each data point is averaged from 5000 independent realizations. 

\item Figure  \ref{distributions}, \ref{structures}, and \ref{chi_si}: the simulation is the same as Figure \ref{rdelete}(A, B). The model without resource extinction simulated with eqs. (\ref{Ma1}),  and each data point is averaged from 5000 independent realizations.. The model with resource extinction is simulated with eqs. (\ref{Ma}),  and each data point is averaged from 4000 independent realizations. 

\end{itemize}

\section{Distinction between extinct and surviving species}
In the main text, we show that the value of species packing $\frac{S^*}{M}$ in Fig. \ref{heatmap} and  Fig. \ref{phase}. However, in numerical simulations,  even for the extinct  species, the abundance is never exactly equal 0 due to numerical errors.  As a result, we must choose a threshold to distinguish extinct and surviving species in order to calculate $S^*$. Since we are using the equivalence with convex optimization to solve the generalized consumer-resource models\cite{marsland2019minimum,mehta2019constrained}, we can easily choose a reasonable threshold (e.g. $10^{-10}$ for both species since the surviving species are well separated in two peaks (see Fig. \ref{Nsim}).

\section{Susceptibility matrix for linear resource dynamics}\label{linearSusceptibilities}
In the quasi-static limit , eqs. \ref{Ma2} becomes
$$\frac{dN_i}{dt}=N_i\left(\frac{N_i \sum_\alpha C_{i\alpha}K_\alpha}{ 1 +\sum_j C_{j\alpha}N_j}-m_i \right),$$
which can not be reduced to the Lotka–Volterra model.Therefore, we have to rederive the susceptibility matrix for eqs. \ref{Ma2}.

In order to have a well defined susceptibilities, we introduce an auxiliary variable  $w_\alpha$ and eqs. \ref{Ma2} become
\begin{eqnarray}
\begin{cases}
\frac{d N_i}{dt}  = N_{i}(\sum_{\beta} \bar{C}_{i \beta} R_{\beta} -  m_i)\\
\frac{d R_\alpha}{dt}    = \kappa_{\alpha}  - w_\alpha R_\alpha -  \sum_{j}N_{j} \bar{C}_{j\alpha}R_\alpha.
\label{Ma3}
\end{cases}
\end{eqnarray}
At the end we can set $w_\alpha$  to 1 to recover eqs. \ref{Ma2}. Employing the results from \cite{cui2019effect} and \cite{cui2020perturbative},  the new susceptibility matrix is
\begin{eqnarray}
\chi^R_{\alpha \beta}=-\frac{\partial \bar{R}_\alpha}{\partial \omega_\beta},
\chi^N_{i \alpha}=-\frac{\partial \bar{N}_i}{\partial \omega_\alpha}, 
\nu^R_{\alpha i}=\frac{\partial \bar{R}_\alpha }{\partial m_i}, 
\nu^N_{i j}=\frac{\partial \bar{N}_i }{\partial m_j}
\end{eqnarray}
where the bar $\bar{X}$ over the variable $X$  denotes the steady-state (equilibrium) solution. 

For the extinct species and resources, by definition the susceptibilities are zero. For this reason, we focus only on the surviving resources and species.
At steady-state, eqs. \ref{Ma3} gives:
\begin{eqnarray}
0&=&\sum_{\alpha\in \mathbf{M}} C_{i \alpha} \bar{R}_{ \alpha} - m_i\\
0&=&K_{\alpha} -\omega_\alpha \bar{R}_\alpha-  \bar{R}_\alpha\sum_{j\in \mathbf{S^*}}\bar{N}_{j} C_{j\alpha} 
\end{eqnarray}
where $\mathbf{S^*}$  denote the sets of species, respectively, that survive in the ecosystem at steady-state and  $\mathbf{M}$ denotes the full sets of resources as they all are nonzero for the linear resource dynamics. Differentiating
these equations yields the relations
\begin{eqnarray} 
0\!=\! \sum_{\alpha\in \mathbf{M}} \!C_{i \alpha} \frac{\partial \bar{R}_\alpha}{\partial \omega_\beta},  &\quad& -\bar{R}_\alpha\delta_{\alpha \beta}\!=\!(\omega_\alpha\!+\!\sum_{j\in \mathbf{S^*}}\bar{N}_{j} C_{j\alpha} )\frac{\partial  \bar{R}_\alpha}{\partial \omega_\beta}\!+\! \sum_{j\in \mathbf{S^*}}\frac{\partial  \bar{N}_j}{\partial \omega_\beta} C_{j\alpha} \bar{R}_\alpha\nonumber \\
\delta_{i j}\!=\! \sum_{\alpha\in \mathbf{M}}\! C_{i \alpha} \frac{\partial  \bar{R}_\alpha}{\partial m_j},  &\quad&
0\!=\!(\omega_\alpha\!+\!\sum_{j\in \mathbf{S^*}}\bar{N}_{j} C_{j\alpha} )\frac{\partial  \bar{R}_\alpha}{\partial m_i}\!+ \!\sum_{j\in \mathbf{S^*}}\frac{\partial  \bar{N}_j}{\partial m_i} C_{j\alpha}\bar{R}_\alpha.
\end{eqnarray} 
Substituting in for the partial derivatives using the susceptibility matrices defined above, we have:
\begin{eqnarray} 
0=\sum_{\alpha\in \mathbf{M}} C_{i \alpha} \chi^R_{\alpha \beta},&\quad&
\bar{R}_\alpha\delta_{\alpha \beta}=(\omega_\alpha\!+\!\sum_{j\in \mathbf{S^*}}\bar{N}_{j} C_{j\alpha} )\chi^R_{\alpha \beta}+ \sum_{j\in \mathbf{S^*}}\chi^N_{j \beta} C_{j\alpha}\bar{R}_\alpha \nonumber \\
\delta_{i j}=\sum_{\alpha\in \mathbf{M}} C_{i \alpha} \nu^R_{\alpha j}, &\quad&
0= (\omega_\alpha\!+\!\sum_{j\in \mathbf{S^*}}\bar{N}_{j} C_{j\alpha} )\nu^R_{\alpha i}+ \sum_{j\in \mathbf{S^*}} \nu^N_{j i} C_{j\alpha}\bar{R}_\alpha.
\end{eqnarray} 
These two equations can be written as a single matrix equation for block matrices:
\begin{eqnarray}
\begin{bmatrix}
\mathbf{C}& 0\\\
\text{diag}(W_\alpha) & \mathbf{G}^T
\end{bmatrix}  
\begin{bmatrix}
\mathbf{\nu}^R & \mathbf{\chi}^R\\\
\mathbf{\nu}^N& \mathbf{\chi}^N
\end{bmatrix} =\begin{bmatrix}
\mathds{1} & 0\\\
0& \text{diag}(\bar{R}_\alpha)
\end{bmatrix} 
\end{eqnarray}
where $W_\alpha=\omega_\alpha\!+\!\sum_{j\in \mathbf{S^*}}\bar{N}_{j} C_{j\alpha}$, $G_{i\alpha}=C_{i\alpha}\bar{R}_\alpha$ and diag is the operator transforming a vector to a diagonal matrix.

To solve this equation, we define two $S^* \times S^* $ matrices: $A_{ij}=\sum_{\alpha\in \mathbf{M}^*}C_{i\alpha}C^{T}_{j\alpha}$ and $H_{ij}=(\sum_{\alpha\in \mathbf{M}^*}\frac{\bar{R}_\alpha}{W_{\alpha}}C_{i\alpha}C^T_{j\alpha})^{-1}$. Employing eq. (3.2) for  the square off-diagonal partition, a straightforward calculation yields
\begin{eqnarray}
\renewcommand\arraystretch{2}
\begin{bmatrix}
\mathbf{C}& 0\\\
\text{diag}(W_\alpha) & \mathbf{G}^T
\end{bmatrix}^{-1}
=
\begin{bmatrix}
\sum_{i\in \mathbf{S^*}} \frac{\bar{R}_\alpha}{W_\alpha}C^T_{i\alpha}H_{ij}& \frac{\delta_{\alpha\beta}}{W_{\alpha}}-\sum_{i,j \in \mathbf{S}^*}\frac{\bar{R}_\alpha C^T_{i\alpha}}{W_\alpha}H_{ij}\frac{C_{j\beta}}{W_{\beta}}\\\
-\mathbf{H} & \sum_{j\in \mathbf{S^*}}H_{ij}C_{j\alpha}/W_{\alpha}
\end{bmatrix}
\end{eqnarray}
\begin{eqnarray}
\chi^R_{\alpha \beta}&=&\frac{\bar{R}_\alpha}{W_{\alpha}}\delta_{\alpha\beta}-\sum_{i,j \in \mathbf{S}^*}\frac{\bar{R}_\alpha C^T_{i\alpha}}{W_\alpha}H_{ij}\frac{C_{j\beta}\bar{R}_\beta}{W_{\beta}}\\
\chi^N_{i \alpha}&=&\sum_{j\in \mathbf{S^*}}H_{ij}\frac{C_{j\alpha}\bar{R}_\alpha}{W_{\alpha}}\label{chi1}, 
\quad \nu^R_{\alpha i}=\sum_{j\in \mathbf{S^*}} \frac{\bar{R}_\alpha C^T_{j\alpha}}{W_\alpha}H_{ji}\\
\nu^N_{i j}&=&-H_{ij},\quad  i,j\in \mathbf{S}^* \text{ and } \alpha,\beta\in \mathbf{M}^* \label{nulinear}
\end{eqnarray}


\begin{figure}
	\centering
	\includegraphics[width=0.8\textwidth]{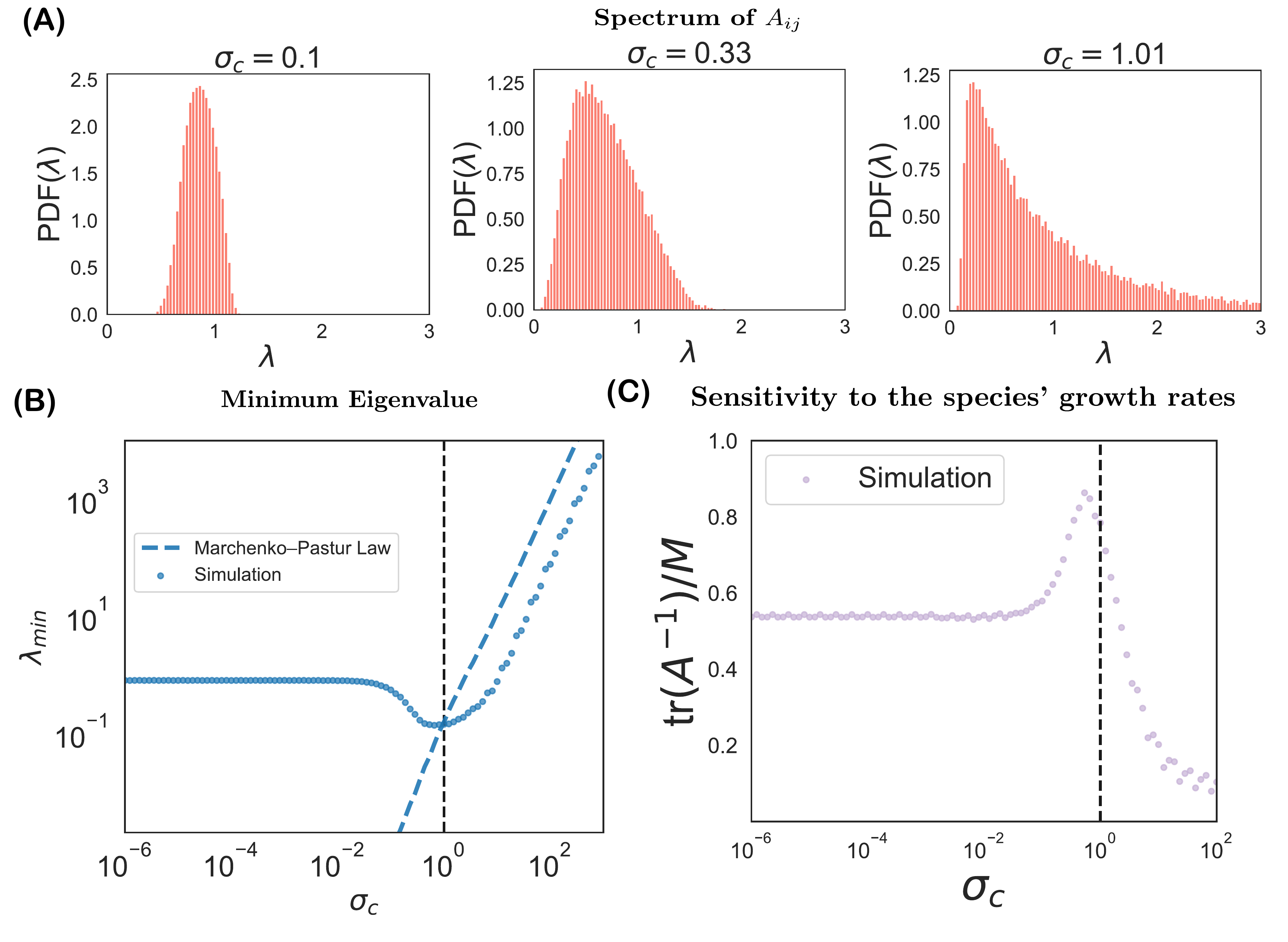}
	\caption{ Reproduce Fig. \ref{pdf} in the main text with model eqs. \ref{Ma1}. The parameters are the same as Fig. \ref{dynamics}.}
	\label{linear}
\end{figure}

We can see that the new susceptibilitie: $H_{ij}=(\sum_{\alpha\in \mathbf{M}^*}\frac{\bar{R}_\alpha}{W_{\alpha}}C_{i\alpha}C^T_{j\alpha})^{-1}$ is different with $A_{ij}=\sum_{\alpha\in \mathbf{M}^*}C_{i\alpha}C^{T}_{j\alpha}$ in eq. \ref{Aij-PM}.  Therefore, it can not behave exactly like Marchenko Pastur distribution, shown in Fig. \ref{linear} (B).  However,  since it is very similar to the Wishart matrix, most of our results are still preserved with eqs. (\ref{Ma1}).

\break
\clearpage 
\newpage
\clearpage

\section{Additional figures}\label{sec:additionfigures}

\begin{figure}[h]
	\centering
	\includegraphics[width=0.4\textwidth]{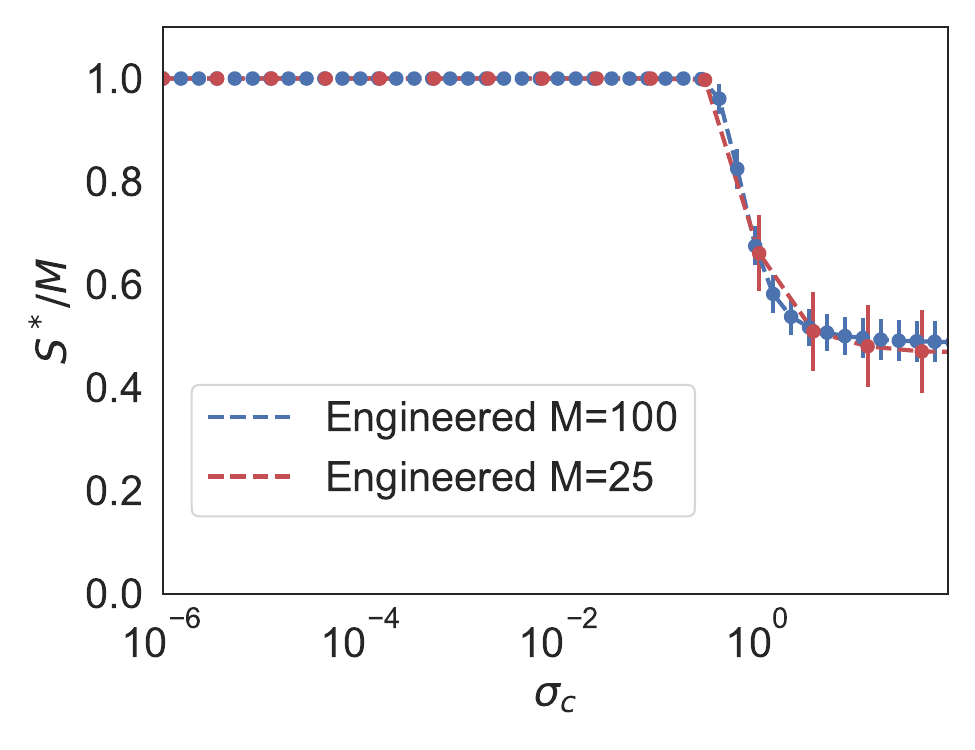}
	\caption{ Reproduce Fig \ref{heatmap} (B): the fraction of surviving species $S^*/M$ vs. $\sigma_c$ for $M = 25$ and $M=100$. It shows $M=25$ is enough to reproduce our result in the main text. Theoretically, numeric converges to our analytical result at the rate of $\frac{1}{M}$.  And it is true that a smaller value of $M$ can result in a larger fluctuation, corresponding to a larger error bar. But the average converges to the same value which is the thermodynamic limit we care about.}
	\label{Msize}
\end{figure}

\end{document}